\documentclass[aps,amssymb,amsmath,amsfonts,twocolumn,pra]{revtex4}
\usepackage[latin1]{inputenc}
\usepackage{amsmath}
\usepackage{graphicx} 
\usepackage{subfigure} 
\usepackage{amsfonts}
\usepackage{hyperref} 
\usepackage{xcolor}
\usepackage{verbatim} 
\usepackage{bbm}
\newcommand{\qed}{\hfill \mbox{\raggedright \rule{.07in}{.1in}}}

\newtheorem{prop}{PROPOSITION}
\newtheorem{teo}{THEOREM}
\newtheorem{corol}{COROLLARY}

\begin{document}
\title{Genuinely multipartite entangled states and orthogonal arrays}
\author{Dardo Goyeneche}
\email{dgoyeneche@cefop.udec.cl}
\affiliation{Departamento de Fis\'{i}ca, Universidad de Concepci\'{o}n, Casilla 160-C, Concepci\'{o}n, Chile\\Center for Optics and Photonics, Universidad de Concepci\'{o}n, Casilla 4012, Concepci\'{o}n, Chile}\vspace{1cm}
\author{Karol {\.Z}yczkowski}
\affiliation{Institute of Physics, Jagiellonian University, ul. Reymonta 4, 30-059 Krak\'{o}w, Poland\\Center for Theoretical Physics, Polish Academy of Sciences, al. Lotnik\'{o}w 32/46, 02-668 Warszawa, Poland}

\date{July 17, 2014}

\begin{abstract}
A pure quantum state of $N$ subsystems with $d$ levels each is called
$k$--multipartite maximally entangled state,
written $k$--uniform, if all its reductions to $k$ qudits are maximally mixed.
These states form a natural generalization of $N$--qudits GHZ states
which belong to the class $1$--uniform states.
We establish a link between the combinatorial notion of orthogonal arrays
and $k$--uniform states and prove the existence of several new
classes of such states for $N$--qudit systems.
In particular, known Hadamard matrices allow us to explicitly
construct $2$--uniform states for an arbitrary number of $N>5$  qubits.
We show that finding a different class of $2$--uniform states
would imply the Hadamard conjecture,
so the full classification of $2$--uniform states seems to be currently out of reach. Furthermore, we establish links between existence of $k$--uniform states, classical and quantum error correction codes and provide a novel graph representation for such states.
\end{abstract}
\maketitle

\section{Introduction}

A multipartite pure state of $N$ qudits is called
entangled if it cannot be written as the tensor product of $N$ single--qudit pure states. There exist also states for which
at least some subsystems can be factorized.
Hence one distinguishes the bi-separable states,
for which one can find a splitting with respect to which
a given state is separable.
A state is called {\sl genuinely entangled} if
all subsystems are correlated and the state is not separable
with respect to any possible splitting of $N$ subsystems \cite{CKW00,HHHH09}.

In the case of multi-partite systems
there exists several different classes of entangled states.
In the simplest case of three qubits there are two non-comparable classes
of entanglement, called GHZ and W, which are not equivalent with respect
to local operations and classical communication \cite{DVC00}.
In the case of $N=4$ qubits the number of inequivalent classes of pure states
grows to nine \cite{VDMV02} or more \cite{CD07},
if another classification is used.
In general the number of parameters describing entanglement
grows exponentially with the number $N$ of subsystems \cite{LPS99}.

Investigation of highly entangled states of several qubits,
initiated by Gisin and Bechmann-Pasquinucci \cite{GBP98},
was then continued in the specific case of four qubits by
Higuchi and Sudbery \cite{HS00}.
Later on the issue of identifying multipartite pure states
which are distinguished by maximizing certain
measures of entanglement was analyzed in
\cite{KNM02,BSSB05,OS06,Borras,BH07,AFOV08} and was
 further developed in \cite{DO09,MGP10,TST10}.

In the case of bipartite systems, one distinguishes the Bell states
and their generalizations for two $d$--level systems,
for which reduced states are maximally mixed.
A class of multi--qubit pure states with a stronger property,
that every possible one-qubit reduction is maximally mixed was
analyzed in \cite{GBP98,OS06,FFPP08,FFMPP10} and called perfect maximally multipartite entangled state.
A further important contribution on these issues was provided recently
by Arnaud and Cerf \cite{Arnaud},
who used the name of $k$--{\sl multipartite maximally entangled} pure states.
In the present work we shall call these states {\sl $k$--uniform},
following an earlier paper of Scott  \cite{Scott}. These
states are distinguished by the fact that their reduction
with respect to an arbitrary splitting leaves a $k$-partite
reduced state maximally mixed, so the mean purity of the
reduction, averaged over different choices of the ancillary system,
is minimal \cite{FFMPP10b}.
 In particular, $1$--uniform states are also {\sl balanced}, which means that
the total number each of the $d$ levels appears in
the representation of the state is equal \cite{CKW00,OS10}.

Construction of genuinely multipartite entangled states is an
important open problem in the theory of quantum information,
as these states have numerous applications to
 quantum teleportation, quantum key distribution, dense
 coding and error correcting codes \cite{Scott,Arnaud}.
The main goal of this work is to establish a link between
orthogonal arrays and $k$--uniform states and to make use of it to construct new families
of such states for an arbitrary number of subsystems.

This paper is organized as follows: In Section II we introduce $k$--uniform states and resume the \emph{state of the art}.
 In Section III we present basic concepts of orthogonal arrays useful in this work. In Section IV we
 connect $k$--uniform states with orthogonal arrays in a natural way.
This allows us to prove the existence of new families of $k$--uniform states for $N$ subsystems of $d$ levels each.
In Section V we present examples of $k$--uniform states obtained from orthogonal arrays and we study the minimal number of terms required to construct a $k$--uniform state of $N$ qubits.
In Section VI we show that the problem of classifying $2$--uniform states for qubits contains the Hadamard conjecture.
Consequently, the complete classification of $2$--uniform states seems to be currently out of reach.
Fortunately, a different class of 2--uniform states is possible,
so making use of the well--known Hadamard matrices of order $2^m$ and $3\times2^m$
we construct such states for every $N$. The paper is concluded in Section VII,
in which a list of open questions is presented.
Explicitly constructed families of $2$--uniform states of 6 to 15 qubits are listed in Appendix A.
In Appendix B we present further examples of $k$--uniform states of qudits having $d>2$ levels
related to orthogonal arrays. In Appendix C we present an extended
construction of further families of $k$--uniform states.

\section{$\mathbf{k}$--multipartite maximally entangled pure states}
In this section we introduce the concept of $k$--uniform states
\cite{FFPP08,Arnaud} and present a short review of the state of the art.
A pure quantum state of $N$ subsystems with $d$ levels each is called $k$--maximally entangled,
 written $k$--uniform, if every reduction to $k$ qudits is maximally mixed. For example, GHZ states belong to the class $1$--uniform and W states do not belong to any class of $k$--uniform states. The
general problem to construct $k$--uniform states is difficult and explicit solutions are
known for systems consisting of a few subsystems only. For convenience of the
reader we present here some of the most important results known in the literature.

We start recalling the notion of local equivalence.
Two pure states $|\psi\rangle$ and $|\phi\rangle$
of a bi--partite system are called {\sl locally equivalent},
written $|\psi\rangle \sim_{\rm loc} |\phi\rangle$,
if there exists a product unitary matrix $U_A \otimes U_B$
such that $|\psi\rangle = U_A \otimes U_B |\phi\rangle$.
In the case of $N$--partite systems the product unitary matrix
factorizes into a tensor product of $N$ unitaries.
By definition, if a given state $|\Psi\rangle$
belongs to the class of $k$--uniform states, so does any other
locally equivalent state.

In the case of two qubits one distinguishes four mutually
orthogonal, maximally entangled {\sl Bell states},
\begin{equation}
\label{Bell2}
|\Phi^{\pm}_{2}\rangle=|00\rangle\pm|11\rangle\mbox{\hspace{0.3cm}and\hspace{0.3cm}}|\Psi^{\pm}_2\rangle=|01\rangle\pm|10\rangle .
\end{equation}
These locally equivalent states are $1$--uniform. Moreover, it is easy to show
that any state of this class has to be locally equivalent to the Bell state
$|\Phi^{+}_{2}\rangle$.

For three qubits the only $1$--uniform state is the GHZ state \cite{GHZ}
\begin{equation}
\label{GHZ3}
|GHZ\rangle=|000\rangle+|111\rangle,
\end{equation}
up a local unitary transformation. For brevity we shall omit in the paper
the normalization factors, provided they are the same for each term.

 A class of $1$--uniform states of $N$ qudits systems of $d$ levels is easy to find
\begin{equation}
|GHZ_{N}^d \rangle=\sum_{i=0}^{d-1}|i\rangle^{\otimes N}.
\label{GHZn}
\end{equation}

For three qubits there exists another class of maximally entangled states
called W,
\begin{equation}
\label{W3}
|W\rangle=|100\rangle+|010\rangle+|001\rangle .
\end{equation}
However,  the reductions to $1$--qubit systems are not maximally mixed, so
$|W\rangle$ is neither  a $1$--uniform state
nor a balanced state \cite{OS10}.

In the case of four qubits it is straightforward to write some $1$--uniform states.
Any state locally equivalent to the four--qubit state $|GHZ_4^2\rangle$
 -- a special case of (\ref{GHZn}) -- does the job. Another example reads
\begin{eqnarray}
|\Phi_4\rangle&=& |0000\rangle+|0011\rangle+|0101\rangle+|0110
\rangle+\nonumber\\&&|1001\rangle+|1010\rangle+|1100\rangle+|1111\rangle.\nonumber\\
\end{eqnarray}
This state is locally equivalent to the 1-uniform state of four qubits $|HS\rangle$ \cite{HS00}. It is worth to emphasize that there exist no $2$--uniform
states of four qubits \cite{HS00,Borras,BH07,FFPP08,Gour}.
This fact was interpreted as a symptom of {\sl frustration}  \cite{FFMPP10b},
as the requirements that the entanglement is maximal for all possible bipartitions
of the system become conflicting. The phenomenon of frustration is known in
various matrix models including spin glasses \cite{CKPR95}, for which
 a phase transition takes place.
 In the case of $N=4$ qubit systems, or other systems for which
uniform states do not exist, it is interesting to identify
special states, for which the average entanglement is maximal.
More formally, one looks for states for which the mean purity of the $k$-partite
reduced state, averaged over different choices of the ancillary system,
is minimal \cite{FFMPP10b,ZYZ13}. However, it has been proven that such minimization procedure do not help to uniquely identify maximally entangled states when a $k$-uniform state does not exist \cite{Borras2}. For five qubits there are $2$--uniform states:
\begin{eqnarray}\label{Phi5first}
|\Phi_{5}\rangle&=&-|00000\rangle+|01111\rangle-|10011\rangle+\nonumber\\&&|11100\rangle+|00110\rangle+|01001\rangle+\nonumber\\&&|10101\rangle+|11010\rangle.
\end{eqnarray}

This state,  also called $|0_L\rangle$,  has been used to distribute quantum information
over five qubits \cite{Laflamme}. It is worth noting that a $k$--uniform state is also $k^{\prime}$--uniform
 state with $0<k^{\prime}<k$. Thus, the state $|\Phi_{5}\rangle$ is also $1$--uniform.
It is easy to see that there is no $3$--uniform state of 5 qubits,
as  a $k$--uniform state of $N$ qubits can exists if \cite{Scott}
\begin{equation}
\label{Scott}
k\leq\lfloor N/2\rfloor,
\end{equation}
where $\lfloor\cdot\rfloor$ denotes integer part.
If this bound is not satisfied then the dimension of the reduced state
is larger than the size of the ancillary space, so the reduction
of a pure state cannot be maximally mixed.
Observe that for $N=2,3$ and $5$ the above inequality is saturated,
however for $N=4$ it is not the case.

 Additionally, there exist  $3$--uniform states of six qubits \cite{Borras}:
\begin{eqnarray}
\label{Phi6}
|\Phi_{6}\rangle&=&-|000000\rangle+|001111\rangle-|010011\rangle+\nonumber\\
&&|011100\rangle+|000110\rangle+|001001\rangle+\nonumber\\
&&|010101\rangle+|011010\rangle-|111111\rangle+\nonumber\\
&&|110000\rangle+|101100\rangle-|100011\rangle+\nonumber\\
&&|111001\rangle+|110110\rangle-|101010\rangle-\nonumber\\
&&|100101\rangle.
\end{eqnarray}
The above examples have been constructed at hand or from computing algorithms. As far as we know, the first expression of a 3-uniform state of six qubits was found by Borras \emph{et al}. \cite{Borras}.
In general, for $N>6$ the existence of $k$--uniform states for $k>1$ remains open.

In this work we are going to use the combinatorial notion of \emph{orthogonal arrays}
and demonstrate that they form a suitable tool for studying genuine multipartite entangled pure states.

\section{Orthogonal arrays}
Orthogonal arrays are combinatorial arrangements introduced by Rao \cite{Rao1} in 1946.
 They have a close connection to codes, error correcting codes, difference schemes, Latin squares and
 Hadamard matrices -- see \cite{Hedayat}. The most important applications of orthogonal arrays are
given in statistics and design of experiments.
An $r\times N$ array $A$ with entries taken from a set $S$ with $d$ elements
 is said to be an {\sl orthogonal array} with $r$ runs, $N$ factors,
 $d$ levels, strength $k$ and index $\lambda$
if every $r\times k$ subarray of $A$ contains each $k-$tuple of symbols from $S$
 exactly $\lambda$ times as a row.
Here, $r$ and $N$ denote the number of rows and columns of $A$, respectively, while $d$ is the cardinality of the set $S$. That is, the level $d$ is the number of different symbols appearing in $A$.

An excellent introduction to orthogonal arrays and their applications is provided
in the book of Hedayat, Sloane and Stufken \cite{Hedayat}.
Furthermore, extensive catalogs of orthogonal arrays can be found in
the handbook \cite{CD96} and in the websites
of Sloane \cite{Sloane} and Kuhfeld \cite{Kuhfeld}.
The ordering of parameters characterizing orthogonal arrays used in this work follows
the standard notation introduced by Rao \cite{Rao1}:
\begin{equation*}
\mbox{OA($r,N,d,k$)}.
\end{equation*}
However, the labels of the parameters has been conveniently adapted here to quantum theory.
The same ordering is used in \cite{Hedayat,Sloane,Kuhfeld}, but since some other conventions
are also present in the literature, the reader is advised to check the order of the parameters
before using a given resource on orthogonal arrays.

\begin{figure}[!h]
\centering 
{\includegraphics[width=6cm]{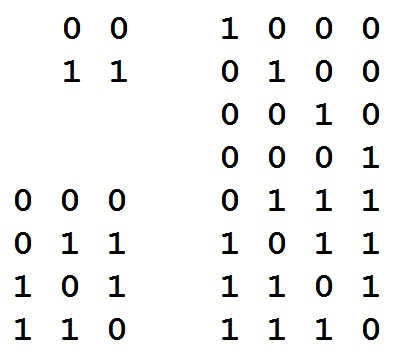}} 
\caption{Orthogonal arrays of strength one (upper left), two (down left) and three (right).
 Their symbolic expressions are OA(2,2,2,1), OA(4,3,2,2) and OA(8,4,2,3), respectively.}
\label{Fig1}
\end{figure}

One usually determines an orthogonal array by the following
four independent parameters $r,N,d$ and $k$,
 while the index $\lambda$ satisfies the relation:
\begin{equation}
\label{lambda}
r=\lambda d^k.
\end{equation}
Additionally, the following {\sl  Rao bounds} hold \cite{Rao2},
\begin{eqnarray}\label{raobounds}
r&\geq&\sum_{i=0}^{k/2}\binom{N}{i}(d-1)^i\hspace{0.5cm}\mbox{if $k$ is even,}\\
r&\geq&\sum_{i=0}^{\frac{k-1}{2}}\binom{N}{i}(d-1)^i+\nonumber\\&&\binom{N-1}{\frac{k-1}{2}}(d-1)^{\frac{k-1}{2}}\hspace{0.5cm}\mbox{if $k$ is odd.}\nonumber\\
\end{eqnarray}
If the parameters are such that the above inequalities are saturated then the
OA is called \emph{tight}. It is easy to check that the examples presented in Fig.\ref{Fig1}
satisfy this property.
 Given an OA($r,N,d,k$) we can easily construct an OA($r,N^{\prime},d,k$) for any $k\leq N^{\prime}\leq N$ by removing from the array $N-N^{\prime}$ columns.
 Therefore, it is interesting to determine the maximal factor $N$ such that an OA($r,N,d,k$)
exists for $r,d$ and $k$ fixed. This maximal factor is defined as
\begin{equation}
N_{max}=f(r,d,k).
\end{equation}
Orthogonal arrays are very useful to design fractional factorial experiments \cite{Rao2}. The parameter $r$ determines the number of \emph{runs} of the experiments. Thus, it is
important to minimize the number $r$ without changing the rest of the parameters defining the experiment. Consequently, we define
\begin{equation}
r_{min}=F(N,d,k),
\end{equation}
as the lowest value of $r$ such that an OA($r,N,d,k$) exists for $N,d$ and $k$ fixed.
Interestingly, the following relationships are satisfied \cite{Hedayat}:
\begin{eqnarray}
F(N,d,k)&=&\min\{r: f(r,d,k)\geq N\},\nonumber\\
f(r,d,k)&\leq&\max\{N: F(N,d,k)\leq r\}.\nonumber
\end{eqnarray}

In what follows we present one of the first important results on the existence of orthogonal arrays, discovered in 1946 by Rao \cite{Rao1}:
\begin{teo}[Rao]\label{TeoRao}
If $d$ is a prime power then an OA($d^n,(d^n-1)/(d-1),d,2$) exists whenever $n\geq2$.
\end{teo}
These OA can be constructed \cite{Rao2} by using Galois fields
but this is not simple to do for high values of $d$.

\section{\emph{\textbf{k}}--uniform quantum states from orthogonal arrays}
\label{OAandKMM}
In this section we study a close connection between orthogonal arrays and
$k$--uniform  quantum states. Let us start by defining a pure state $|\Phi\rangle$
of a system consisting of $N$ qudits,
\begin{equation}
|\Phi\rangle=\sum_{s_1,\dots,s_N}a_{s_1,\dots,s_N}|s_1,\dots,s_N\rangle,
\end{equation}
 where $a_{s_1,\dots,s_N}\in\mathbb{C}$, $s_1,\dots,s_N\in S$ and $S=\{0,\dots,d-1\}$.
The set of vectors $\{|s_1,\dots,s_N\rangle\}$ forms an orthonormal basis of $\mathbb{C}^{d^N}$.
 In fact, this is the \emph{canonical} or \emph{computational} basis.
For simplicity the range of the sums is omitted,
but all of them go from $0$ to $d-1$.
 The density matrix $\rho$  associated to this pure state reads
\begin{eqnarray}\label{pure}
\rho=|\Phi\rangle\langle\Phi|&=&\sum_{s_1,\dots,s_N\atop s^{\prime}_1,\dots,s^{\prime}_N}a_{s_1,\dots,s_N}
a^*_{s^{\prime}_1,\dots,s^{\prime}_N}\nonumber\times\\&&
|s_1,\dots,s_N\rangle\langle s^{\prime}_1,\dots,s^{\prime}_N|.\nonumber\\
\end{eqnarray}
Let us divide the system $\mathcal{S}$ in two parts $\mathcal{S}_A$ and $\mathcal{S}_B$ each containing $N_A$ and $N_B$ qudits, respectively such that $N=N_A+N_B$. In order to find the density matrix associated to $\mathcal{S}_A$ we have to consider the reduced state
\begin{equation}
\rho_A=\mathrm{Tr}_B(\rho_{AB}),
\end{equation}
where $\rho_{AB}$ denotes the state (\ref{pure}). Performing the partial trace we get
\begin{widetext}
\begin{eqnarray}
\hspace{-0.3cm}\rho_A&=&\hspace{-0.1cm}\mathrm{Tr}_B\left(\sum_{s_1,\dots,s_N\atop s^{\prime}_1,\dots,s^{\prime}_N}a_{s_1,\dots,s_N}a^*_{s^{\prime}_1,\dots,s^{\prime}_N}|s_1,\dots,s_N\rangle\langle s^{\prime}_1,\dots,s^{\prime}_N|\right) \nonumber \\
&=&\hspace{-0.3cm}\sum_{s_1,\dots,s_N\atop s^{\prime}_1,\dots,s^{\prime}_N}a_{s_1,\dots,s_N}a^*_{s^{\prime}_1,\dots,s^{\prime}_N}\mathrm{Tr}_B\left(|s_1,\dots,s_N\rangle\langle s^{\prime}_1,\dots,s^{\prime}_N|\right) \nonumber \\
&=&\hspace{-0.3cm}\sum_{s_1,\dots,s_N\atop s^{\prime}_1,\dots,s^{\prime}_N}a_{s_1,\dots,s_N}a^*_{s^{\prime}_1,\dots,s^{\prime}_N}\,
\langle s^{\prime}_{N_A+1},\dots,s^{\prime}_N|s_{N_A+1},\dots
,s_N\rangle\,|s_1,\dots,s_{N_A}\rangle\langle s^{\prime}_1,\dots,s^{\prime}_{N_A}| \label{rhoA}.
\end{eqnarray}
\end{widetext}
The following reasoning deserves a special attention as it contains the key point of our work.
First, let us assume that every coefficient $a_{s_1,\dots,s_N}$ is \emph{zero} or \emph{one} for simplicity.
 Therefore, the state $|\Phi\rangle$ can be written as a superposition of $r$ product states,
\begin{equation}\label{pure2}
|\Phi\rangle=|s^1_1,s^1_2,\dots,s^1_N\rangle+\dots+|s^r_1,s^r_2,\dots,s^r_N\rangle,
\end{equation}
where the upper index $i$ on $s$ denotes the $i-th$ term in the linear decomposition of $|\Phi\rangle$.
Secondly, let us now arrange the symbols appearing in Eq.(\ref{pure2}) in an array as follows
\begin{equation}\label{array}
\begin{array}{cccc}
    s^1_1 & s^1_2 & \dots & s^1_N \\
    s^2_1 & s^2_2 & \dots & s^2_N \\
    \vdots & \vdots & \dots& \vdots \\
    s^r_1 & s^r_2 & \dots & s^r_N
  \end{array}.
\end{equation}
So, every column of the array is identified with a particular qudit and every row corresponds to a linear term of the state. Here, we are interested to study pure states having maximally mixed reductions. This means that the reduced system $\rho_A$ must be proportional to the identity matrix, independently of the number $N_B$ of qudits traced out. By imposing this requirement into  Eq.(\ref{rhoA}) and considering Eqs.(\ref{pure2}) and (\ref{array}) we find two basic ingredients allowing to construct $k$--uniform states:
\begin{itemize}\label{AB}
\item[\textbf{\emph{(A)}}] \textbf{\emph{Uniformity:} }

The sequence of $N_A$ symbols appearing in every row of every subset of $N_A$ columns of the array given in Eq.(\ref{array}) \emph{is repeated} the same number of times. This
implies that the diagonal of the reduction $\rho_A$ is uniform, as all its elements are equal.
\item[\textbf{\emph{(B)}}] \textbf{\emph{Diagonality:}}

The sequence of $N_B$ symbols appearing in every row of a subset of $N_B$ columns \emph{is not repeated} along the $r$ rows. Due to this property
the reduced density matrix $\rho_A$ becomes diagonal.
\end{itemize}

\begin{figure}[!h]
\centering 
{\includegraphics[width=6cm]{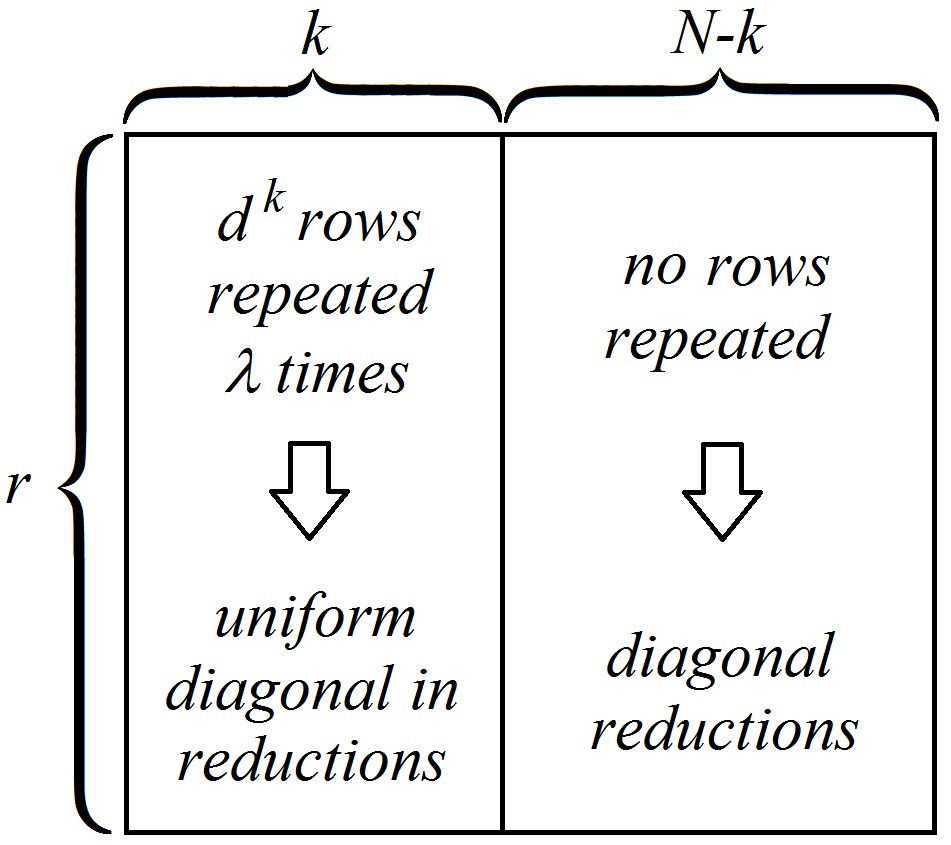}} 
\caption{Connection between orthogonal arrays and $k$--uniform states of $N$ qudits,
   where $k\le N-k$.}
\label{Fig2}
\end{figure}

The above two conditions are sufficient to find $k$--uniform states of $N$ qudits of $d$ levels.
Note that the uniformity condition \emph{(A)} implies that the array defined in Eq.(\ref{array}) is an \emph{orthogonal array} (OA).
From a physics point of view the first condition, concerns the reduced state of the subsystem $A$, while the second one deals with the environment $B$, with respect to which
the partial trace is performed.
 Moreover, there exists a perfect match between the parameters of
an OA($r,N,d,k$) and the parameters of a $k$--uniform state, as we can see in Table I.\vspace{0.5cm}
\begin{widetext}
\begin{center}
\begin{table}[!h]
\begin{tabular}{c|c|c}
\hline \hline
    & orthogonal arrays & multipartite quantum  state $|\Phi\rangle$ \\
  \hline
  \hspace{0.5cm}$r$\hspace{0.5cm} & Runs & Number of linear terms in the state \\
  $N$ & Factors & Number of qudits \\
  $d$ & Levels & dimension of the subsystem ($d=2$ for qubits) \\
  $k$ & \hspace{0.5cm}Strength\hspace{0.5cm} & \hspace{0.5cm} class of entanglement ($k$--uniform)\hspace{0.5cm} \\
\hline \hline
\end{tabular}\label{Tab1}
\caption{Correspondence between parameters of OA and quantum states.}
\end{table}
\end{center}
\end{widetext}
Fig. \ref{Fig2} presents the basic ingredients of the relation between a multipartite entangled state
and an orthogonal array of $N$ columns and $r$ rows.
It will be thus convenient to introduce the following class of arrays.

{\bf Definition}. An orthogonal array OA($r,N,d,k$) is called
 {\sl irredundant}, writted IrOA,
 if removing from the array any $k$ columns
all remaining $r$ rows, containing $N-k$ symbols each, are different.

If this condition is not fulfilled, certain
remaining rows are equal and carry some redundant information.
Any irredundant OA satisfies thus both conditions \emph{(A)} and \emph{(B)}
and allows us to construct  a $k$--uniform state. However,
condition \emph{(B)} is not necessary, as there
exist quantum states, e.g. locally equivalent to states
constructed by an IrOA, which are $k$--uniform
but do not satisfy the diagonality property.

On the other hand, we do not know whether
 every $k$--uniform state is related by a local unitary transformation
to a $k$--uniform state constructed with an irredundant orthogonal array.

Since the seminal work of Rao \cite{Rao1} the theory of orthogonal arrays has been developed significantly. Many important theorems concerning the existence of OA have been found and  direct
connections with other combinatorial arrangements have been established.
We translate here some of these relevant theorems to the construction of $k$--uniform quantum states.
Interestingly, there is a match between basic properties of OA and $k$--uniform states.
For OA the following properties hold --
see pages 4 and 5 in the book of Hedayat \emph{et al.} \cite{Hedayat}:
\begin{itemize}
\item[\emph{(i)}]   The parameters of an orthogonal array satisfy equality $\lambda=r/d^k$.
\item[\emph{(ii)}]  Any OA($r,N,d,k$) is also an OA($r,N,d,k^{\prime}$) for every $k^{\prime}<k$.
\item[\emph{(iii)}] A permutation of the runs or factors in an OA results in an OA with the same parameters.
\item[\emph{(iv)}] A permutation of the levels of any factor in an OA results in an OA with the same parameters.
\item[\emph{(v)}] Any $r\times N^{\prime}$ subarray of an OA($r,N,d,k$) is an OA($r,N^{\prime},d,k^{\prime}$) where $k^{\prime}=\min\{N^{\prime},k\}$ and $N^{\prime}<N$.
\item[\emph{(vi)}] Taking the runs in an OA($r,N,d,k$) that begins with a particular symbol and omitting the first column yields an OA($r/d,N-1,d,k-1$).
\item[\emph{(vii)}] Take $m$ orthogonal arrays  $A_i={\rm OA}(r_i,N,d,k_i$), for $i=0,\dots,m-1$
and define the array $A$ as their juxtaposition.
\begin{equation}\label{juxta}
\left[\begin{array}{c}
A_0\\ A_1 \\ \vdots \\ A_{m-1}
\end{array}\right] .
\end{equation}
Then $A$ is an OA($r,N,d,k$), where $N=N_0+\dots+N_{m-1}$ and the strength is $k$ for some $k\geq\min\{k_0,\dots,k_{m-1}\}$. Furthermore, if $m=d$ and each $A_i$ is an OA($r,N,d,k$), after appending a 0 to each row of $A_0$, a 1 to each row of $A_1$ and so on, we obtain an OA($dr,N+1,d,k$).
\end{itemize}
Now, let us translate the above properties to the setup of $k$--uniform quantum states:
\begin{itemize}
\item[\emph{(i')}] All the reduced density matrices of a $k$--uniform state satisfy $\mathrm{Tr}(\rho)=1$.
\smallskip
\item[\emph{(ii')}] A $k$--uniform state is also a $k^{\prime}$--uniform for every $k^{\prime}<k$.
\smallskip
\item[\emph{(iii')}] A permutation of terms or qudits in a $k$--uniform
 state lead us to a $k$--uniform state.
\smallskip
\item[\emph{(iv')}] Any permutation of the $d$ symbols defining a qudit
(e.g. caused by the Pauli $X$--gate or NOT gate),
 does not change the $k$--uniform property.
\smallskip
\item[\emph{(v')}] Any $k$--uniform state of $N$ qudits can be reduced to a
$k^{\prime}$--uniform state of $N^{\prime}<N$ qudits, where $k^{\prime}=\min\{N^{\prime},k\}$
(whenever the property (B) of p.. \pageref{AB} holds). For example, it fails for $k^{\prime}\leq N^{\prime}/2$).
\smallskip
\item[\emph{(vi')}] We can always decompose a $k$--uniform states as a function of $k^{\prime}$--uniform states with $k^{\prime}\leq k$. For example, a $2$--uniform state of 5 qubits can be decomposed in the following way:
\begin{equation}
|\Psi_5\rangle=|0\rangle|\Psi_{4a}\rangle+|1\rangle|\Psi_{4b}\rangle,
\end{equation}
where
\begin{equation}\label{Phi4a}
|\Psi_{4a}\rangle=|0000\rangle+|1010\rangle+|0101\rangle+|1111\rangle,
\end{equation}
and
\begin{equation}\label{Phi4b}
|\Psi_{4b}\rangle=|0011\rangle+|1001\rangle+|0110\rangle+|1100\rangle,
\end{equation}
are orthogonal $1$--uniform states of 4 qubits.
\smallskip
\item[\emph{(vii')}] If the states $|\Phi_i\rangle,\,i=0,\dots,m-1$ are $k_i$--uniform and come from OA($r_i,N,d,k_i$) then the state $|\Psi\rangle=\sum_{i=0}^{m-1}|\Phi_i\rangle$ is a $k$--uniform state for some $k\geq\min\{k_0,\dots,k_{m-1}\}$. For instance, the sum of the 4 qubits states $|\Psi_{4a}\rangle$ and $|\Psi_{4b}\rangle$ given in Eqs.(\ref{Phi4a}) and (\ref{Phi4b}), respectively, forms a $1$--uniform state.
Furthermore, if $m=d$ and each state $|\Phi_i\rangle$ of $N$ qudits is $k$--uniform then the state
\begin{equation}
\label{phiphii}
|\Phi\rangle=\sum_{i=0}^{m-1}|i\rangle|\Phi_i\rangle,
\end{equation}
is a $k$--uniform state of $N+1$ qudits, whenever the diagonality property (B) holds.
\end{itemize}
The latter condition is crucial, as shown with an example of the following
state of 6 qubits,
\begin{eqnarray}
|\Psi_6\rangle&=&|0\rangle|\Phi_5\rangle+|1\rangle\overline{|\Phi_5\rangle}\nonumber\\
&=&-|000000\rangle+|001111\rangle-|010011\rangle+\nonumber\\
&&|011100\rangle+|000110\rangle+|001001\rangle+\nonumber\\
&&|010101\rangle+|011010\rangle-|100000\rangle+\nonumber\\
&&|101111\rangle-|110011\rangle+|111100\rangle+\nonumber\\
&&|100110\rangle+|101001\rangle+|110101\rangle+\nonumber\\
&&|111010\rangle,
\end{eqnarray}
where the overbar denotes the flip operation performed on every qubit.
In this case the property (B) does not hold, so the state $|\Psi_6\rangle$
is not $2$--uniform nor even $1$--uniform.
However, $3$--uniform states of 6 qubits can be obtained in this way
for an appropriate choice of states -- see Eq.(5) in \cite{Arnaud}.

In decomposition (\ref{phiphii}) every $k_i$--uniform state $|\Phi_i\rangle$
must come from an OA($r_i,N,d,k_i$) in order to get a $k$ uniform state with $k\geq\min\{k_0,\dots,k_{m-1}\}$. For instance, the following $7$--qubit state:
\begin{widetext}
\begin{eqnarray}\label{Phi7}
|\Phi_7\rangle&=&|0\rangle|\Phi_6\rangle+|1\rangle\overline{|\Phi_6\rangle}\nonumber\\
&=&-|0000000\rangle+|0001111\rangle-|0010011\rangle+|0011100\rangle+|0000110\rangle+|0001001\rangle+\nonumber\\
&&|0010101\rangle+|0011010\rangle-|0111111\rangle+|0110000\rangle+|0101100\rangle-|0100011\rangle+\nonumber\\
&&|0111001\rangle+|0110110\rangle-|0101010\rangle-|0100101\rangle-|1111111\rangle+|1110000\rangle-\nonumber\\
&&|1101100\rangle+|1100011\rangle+|1111001\rangle+|1110110\rangle+|1101010\rangle+|1100101\rangle-\nonumber\\
&&|1000000\rangle+|1001111\rangle+|1010011\rangle-|1011100\rangle+|1000110\rangle+|1001001\rangle-\nonumber\\
&&|1010101\rangle-|1011010\rangle,
\end{eqnarray}
\end{widetext}
is constructed from the $3$--uniform state of $6$--qubit $|\Phi_6\rangle$ defined in Eq.(\ref{Phi6}). The state (\ref{Phi7}) is an \emph{almost} $3$--uniform state, as only 3 out
of its $\binom{7}{3}=35$ reductions to three qubits are not maximally mixed.
 These three reductions are identically equal
and have four non--zero eigenvalues equal to $1/4$.
Evidently, three sets of $4$ columns of the orthogonal arrays forming this state have repeated rows and, consequently, the diagonality property \emph{(B)} does not hold. Therefore, the question whether a $3$--uniform state of 7 qubits exists remains open.
Additionally, a $2$--uniform state of 7 qubits having 8 terms appears in Appendix \ref{AppendixA} and another one having 64 terms in Appendix \ref{AppendixB}. In Table \ref{Tabla} we resume the existence of $k$--uniform states for a few number of qubits. Interestingly, some $k$--uniform states seem to require negative terms.

We have shown that orthogonal arrays are useful to construct $k$--uniform states.
However, the relation between these objects is not injective.
The following proposition connects orthogonal arrays with a subset
of multipartite quantum states.

\begin{prop}
For every orthogonal array OA($r,N,d,k$) there exists a quantum state of $N$ qudits
such that every reduction to $k$ qudits has its $d^k$ diagonal entries equal to $d^{-k}$.
\end{prop}

Observe that these pure states are not necessarily entangled.
Moreover, the reductions to $k$ qudits  are not necessarily diagonal.
This proposition can be inferred directly from the scheme presented in Fig.\ref{Fig2}.
Note that the entries of such states must have the same amplitude,
which follows from Eq.(\ref{rhoA}).

\begin{table}
\label{Tabla}
\begin{tabular}{|c|c|c|c|c|c|c|c|}
\hline \hline
$k \,\backslash \ N$ & 2 & 3& 4 & 5& 6 & 7& 8 \\
\hline 1&p&p&p&p&p&p&p\\
\hline 2&-&-&0&n&p&p&p\\
\hline 3&-&-&-&-&n&?&p\\
\hline 4&-&-&-&-&-&-&0\\
\hline \hline
\end{tabular}
\caption{Existence of $k$--uniform states for qubits. Here, $p/n$ denote that a state constructed as a superposition of product states with all positive / some negative coefficients is known in the literature. Symbol "-" denotes that such a state cannot exist, as the
necessary condition $k\le N/2$ is not fulfilled, while $0$ means that such
states do not exist, although the necessary condition is satisfied.}
\end{table}

In order to find a $k$--uniform state we require an irredundant OA satisfying the diagonality property (B). Interestingly, this condition holds for any OA with index unity ($\lambda=1$):
\begin{teo}
\label{OAkMM}
Every OA($r,N,d,k$) of index unity is irredundant
so it is equivalent to a $k$--uniform state of $N$ qudits whenever $k\leq N/2$.
\end{teo}
Proof: Given that $\lambda=1$ every $k$--tuple of symbols appears only one time along the rows of the OA. Moreover, all the possible combinations of the $d$ different symbols appear along the $r=d^k$ rows. This means that any set of $k^{\prime}>k$ columns has repeated rows. Thus, to satisfy the irredundancy
property it is enough to assume that $k\leq N-k$,  which means that $k\leq N/2$ -- see Fig.\ref{Fig2}.
The reciprocal implication is straightforward from the definition of $k$--uniform states and OA. \qed

\medskip
We note that OA of index unity contain all possible combinations of symbols in every subset of $k$ columns ($r=d^k$). The bound $k\leq N/2$, firstly discussed by Scott \cite{Scott} for qubits,
 holds in general for subsystems with an arbitrary number of $d$ levels.
As we will show later it is also predicted by the quantum Singleton bound.
In Appendix \ref{AppendixB} we present a list of OA of index unity obtained from the catalog of N. Sloane \cite{Sloane}.
 In general, these OA are not easy to find as one
needs to use tables of Galois fields.
An efficient algorithm to construct these tables is presented in Chapter 4 of the book \cite{Torres}.
Given that OA with index unity are relevant to construct $k$--uniform states let us mention
two important theorems \cite{Bush}:

\begin{teo}[Bush]
If $d\geq2$ is a prime power number then an OA($d^k,d+1,d,k$) of index unity exists whenever $d\geq k-1\geq0$.
\end{teo}
\begin{teo}[Bush]
If $d=2^m$ and $m\geq1$ then there exist an OA($d^3,d+2,d,3$).
\end{teo}
Also, the following Corollary of Theorem 3.7 (see \cite{Hedayat}, p.. 41) is remarkably important here:
\begin{corol}[Hedayat]
If $d=2^m$ and $m\geq1$ then there exists an OA($d^{d-1},d+2,d,d-1$).
\end{corol}

The translation of these three results to the theory of quantum entanglement is resumed in the following proposition:
\begin{prop}\label{Existence}
The following $k$--uniform states exist:
\begin{enumerate}
\item[\emph{(i)}] $k$--uniform states of $d+1$ qudits with $d$ levels, where $d\geq2$ and $k\leq\frac{d+1}{2}$.
\item[\emph{(ii)}] 3--uniform states of $2^m+2$ qudits with $2^m$ levels, where $m\geq2$.
\item[\emph{(iii)}] $2^m-1$--uniform states of $2^m+2$ qudits with $2^m$ levels, where $m=2,4$.
\end{enumerate}
\end{prop}

\begin{table}
\label{Tabla2}
\begin{tabular}{|c|c|c|c|c|c|c|c|}
\hline \hline
$N \ \,\ \backslash \ d$ & 2 & 3& 4 & 5& 6 & 7& 8 \\
\hline 2&\checkmark&\checkmark&\checkmark&\checkmark&\checkmark&\checkmark&\checkmark\\
\hline 4&-&\checkmark&\checkmark&\checkmark&\textbf{?}&\checkmark&\checkmark\\
\hline 6&\checkmark&\checkmark&\checkmark &\checkmark  &\checkmark&\checkmark&\checkmark\\
\hline 8&-&\textbf{?}&\textbf{?}&\textbf{?}&\textbf{?}&\checkmark &\checkmark\\
\hline \hline
\end{tabular}
\caption{Existence of $k$-uniform states of $d$ levels for the highest
strength $k=N/2$. The first row corresponds to bipartite systems and it includes the GHZ states (i.e. $k=1$). Here, the symbol \checkmark\, denotes existing states, '-'\, that they do not exist and \textbf{?} open existence.}
\end{table}

A $k$--uniform state has the maximal attainable value $k=N/2$ for $N$ even.
States attaining this bound are known as \emph{absolutely maximally entangled states} \cite{Helwig} and
their existence is open in general. Such states are remarkably important in quantum information theory; for example, they are equivalent to pure state threshold quantum secret sharing scheme \cite{Helwig2}. As we mentioned, for qubits states the unique solutions of this kind correspond to $1$--uniform states of 2 qubits (Bell states) and 3--uniform states of 6 qubits where, curiously, a 2--uniform state of 4 qubits does not exist. Interestingly, property (i) of the above proposition provides us the existence of an infinite set of absolutely maximally entangled states:
\begin{corol}
\emph{For every $d\geq3$ odd there are $(d+1)/2$--uniform states of $d+1$ qudits with $d$ levels.}
\end{corol}
Also, Properties (ii) and (iii) lead us to the existence of 3--uniform states
of 6 qudits with 4 levels. In Table \ref{Tabla2} we resume the existence
 of $N/2$--uniform states of $N$ qudits with $d$ levels.
Existence of the 3--uniform state of 6 ququarts shown in Table \ref{Tabla2} is based
on the irredundant orthogonal array OA(64,6,4,3) \cite{Sloane}.

\subsection{\emph{\textbf{k}}--uniform states and QECC}
Quantum error correcting codes (QECC) theory deals with the problem of encoding quantum states into qudits such that a small number of errors can be detected, measured and efficiently corrected. QECC are denoted as $((N,K,D))_d$ where $N$ is the length of the code, $K$ is the dimension of the encoding state, $D$ is the Hamming minimum distance and $d$ is the levels number of the qudits system. An introduction to QECC can be found in the recent book of Lidar and Brun \cite{Lidar}. The standard notation used here for QECC (double parenthesis) is in order to avoid confusions with classical codes. A code having a minimum distance $D$ allows to correct an arbitrary number of errors affecting up to $(D-1)/2$ qudits.
It is known that a $((N,K,D))_2$ QECC exists when the quantum Gilbert-Varshamov bound \cite{Ashikhmin} is satisfied:
\begin{equation}\label{GVbound}
\sum_{j=0}^{D-1}3^j\binom{N}{j}\leq2^N/K.
\end{equation}
Note that this inequality seems to be closely related to the Rao bounds given
in Eq.(\ref{raobounds}). However, Rao bounds cannot be used to predict the existence of orthogonal arrays.
Very interestingly, $((N,1,k+1))_d$ QECC are one-to-one connected to $k$--uniform states of $N$ qubits. Thus, following Eq.(\ref{GVbound}) we have:
\begin{equation}
\sum_{j=0}^{k}3^j\binom{N}{j}\leq2^{N-1}.
\label{eq30}
\end{equation}
From this equation we show that $k$--uniform states of $N$ qubits exist for every $k\in\mathbb{N}$ if $N$ is sufficiently large.
However, this inequality does not allows us to find the minimal number $N$
 for which  $k$--uniform states of $N$ qubits do exist.
For instance, (\ref{eq30}) predicts the existence of $3$--uniform states for $N\geq14$ qubits
 and we know that such class of states exists for 6--qubits systems.
Another important inequality for QECC is the quantum Singleton bound \cite{Rains}:
\begin{equation}
N-\log K\geq2(D-1).
\end{equation}
From here we immediately get the upper bound $k\leq N/2$ for any $k$--uniform state of $N$ qudits.
The codes achieving this bound are called {\sl maximal distance separable} (MDS)
and their existence is open in general.  Interestingly, any
violation of this bound would allow us to get perfect copies of quantum states,
which is forbidden by the \emph{no-cloning theorem}.

We have shown above that QECC are partially related to OA through $k$--uniform states.
Furthermore, classical error correcting codes (CECC) are one-to-one connected to OA (see Section 4.3 of \cite{Hedayat}):
\begin{equation}
(N,K,D)_d\,\Leftrightarrow\,OA(K,N,d,k)\,\mbox{for some $k\geq1$}
\end{equation}
where  single parenthesis are used to denote CECC.
For classical codes $(N,K,D)_d$ we also have a version of the Singleton bound:
\begin{equation}
N\leq d^{K-D+1}.
\end{equation}
The codes achieving this bound are the classical
maximal distance separable codes. An interesting result arises here
(see Theorem 4.21, p.. 79 in \cite{Hedayat}):
\begin{prop}\label{MDSCECC}
Classical MDS codes are orthogonal arrays of index unity.
\end{prop}
We have shown, therefore, that classical error correction codes are
also useful to construct $k$--uniform states.
In particular, MDS--CECC are connected with $k$--uniform states if $N=d^{K-D+1}$.
The above proposition also establishes the following connection between classical and quantum
error correction codes:
\begin{prop}
\label{CQECC1}
An MDS--CECC $(N,K,D^{\prime})_d$ is also a QECC $(N,1,D)_d$, for any $K,D^{\prime}$ whenever $D-1\leq N/2$.
\end{prop}
The proof is straightforward from the above discussions. Although classical and quantum error correction codes have been previously related \cite{Grassl},
our results show that some CECC are useful to construct $k$--uniform states
and that an arrangement can define both a CECC and a QECC.

\subsection{Graph representation of \emph{\textbf{k}}--uniform states}\label{Graph}
Graph theory is useful to represent a special kind of
 multipartite maximally entangled pure states known
 as \emph{graph states} \cite{Hein}, construct
quantum error--correcting codes associated with
graphs \cite{SW01,GLG10} or to define
 ensembles of random states with interaction
 between subsystems specified to a graph \cite{CNZ10}.
Here, we show that orthogonal arrays theory
provides a natural graph representation of $k$--uniform states which is not related to the
graph states. Observe that the graph states correspond to quantum systems composed of
subsystems with the same number of levels.
On the other hand, the graph representation applied here is suitable for any number of multi-level qudits systems.
For simplicity   we concentrate in this paper on homogeneous subsystems
and postpone the more general case of heterogeneous subsystems with different number of levels
for a subsequent publication.
 In the following lines we assume basic definitions and properties of graph theory.
Further details can be found in the book  \cite{Bondy} of Bondy and Murty.

\begin{figure}[!h]
\centering 
{\includegraphics[width=6cm]{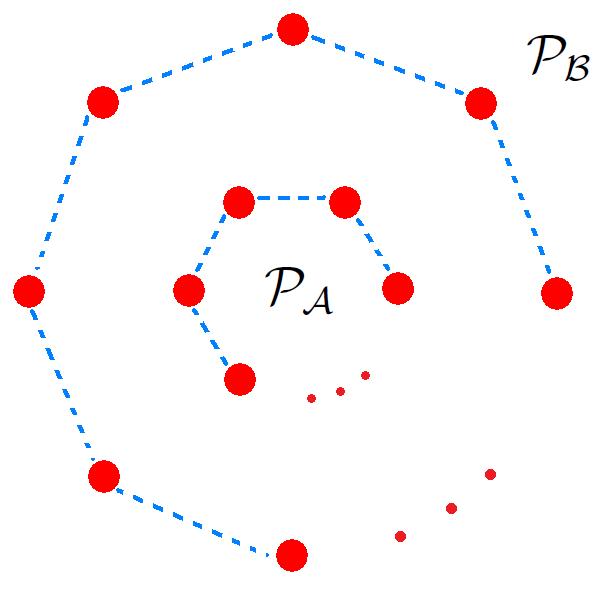}} 
\caption{Graph representation of $k$--uniform states.
The regular polygons $\mathcal{P}_A$ and $\mathcal{P}_B$ represent the subsystems $A$ and $B$, respectively.
The number of vertices of the polygons depend on the number $N$ of $d$--level subsystems, and on the number
 $k\leq k_{max}= \lfloor N/2\rfloor$,
 whereas the edges characterize the degree of entanglement of the state.}
\label{Fig5}
\end{figure}

An orthogonal array OA($r,N,d,k)$ corresponds to a
naturally associated graph composed of $d^k+d^{N-k}$ vertices and $r$ edges.
The vertices define two regular polygons $\mathcal{P}_A$ and $\mathcal{P}_B$,
where $\mathcal{P}_A$ is inscribed into $\mathcal{P}_B$, as shown in Fig.(\ref{Fig5}).

The number of vertices of $\mathcal{P}_A$ and $\mathcal{P}_B$ is $d^k$ and $d^{N-k}$, respectively.
This graph is a sort of graphical representation of a bipartition of the system $AB$,
 where $\mathcal{P}_A$ and $\mathcal{P}_B$ are associated to the subsystems $A$ and $B$.
One has  $\mathcal{H}_{AB}=\mathcal{H}_A\otimes\mathcal{H}_B$ with
 $\mathrm{Dim}(\mathcal{H}_A)=\sharp[\mathcal{P}_A]=d^k$
and $\mathrm{Dim}(\mathcal{H}_B)=\sharp[\mathcal{P}_B]=d^{N-k}$. Note that $\mathcal{P}_A=\mathcal{P}_B$
 if and only if the state is absolutely maximally entangled (i.e. $k=N/2$) \cite{Helwig}.
As we have shown, for $N$--qubit states this is only possible for $N=2$ and $N=6$.

In the following lines we describe our graph construction. Let us define $S=\{0,\dots,d-1\}$ and consider the partition
\begin{equation}\label{partition}
S^{\otimes N}=S_A\oplus S_B,
\end{equation}
such that an OA and its associated pure state are given by:
\begin{equation}
OA=\begin{array}{c}
  s^0_A\hspace{0.2cm}s^0_B \\
  s^1_A\hspace{0.2cm}s^1_B \\
\vdots\\
  s^{r-1}_A\hspace{0.2cm}s^{r-1}_B
\end{array}
\hspace{1cm}|\Phi\rangle=\sum_{i=0}^{r-1}|s^i_A,s^i_B\rangle,
\end{equation}
where $s^i_A\in S_A$ and $s^i_B\in S_B$ for every $i=0,\dots,r-1$
whereas $\mathrm{Dim}(S_A)=k$ and $\mathrm{Dim}(S_B)=N-k$. We assume that every entry of $|\Phi\rangle$ is zero
or one without taking the normalization into account.
Therefore, a simple rule for constructing graphs arises:
\begin{enumerate}
  \item Associate the value $s^i_A\in S_A$ to every vertex of the polygon $\mathcal{P_A}$ and $s^i_B\in S_B$ to every vertex of the polygon $\mathcal{P_B}$, for every $i=0,\dots,r-1$.
  \item Connect the vertex $s^i_A$ to $s^i_B$, for every $i=0,\dots,r-1$.
\end{enumerate}
From here, we note the following interesting properties:
\begin{enumerate}
  \item Polygons $\mathcal{P_A}$ and $\mathcal{P_B}$ are connected by, at least, an edge for entangled states (i.e., if $\mathcal{P_A}$ and $\mathcal{P_B}$ are disconnected then the state is separable).
  \item Two vertices of the same polygon are never connected.
  \item If only one vertex of $\mathcal{P_A}$ (or $\mathcal{P_B}$) is connected then the qudit associated to the corresponding subsystem is not entangled to the rest.
  \item  A  quantum state is $k$--uniform iff the corresponding graph
   satisfies two properties equivalent to

\textbf{\emph{(A') Diagonality:}} every vertex of $\mathcal{P}_B$ is connected, at most, to one edge.

\textbf{\emph{(B') Uniformity:}} every vertex of $\mathcal{P}_A$ is connected to the same number of edges.
\end{enumerate}

\begin{figure}[h!]
\begin{center}
\subfigure[\label{Fig6a}]{
\includegraphics[width=3.5cm]{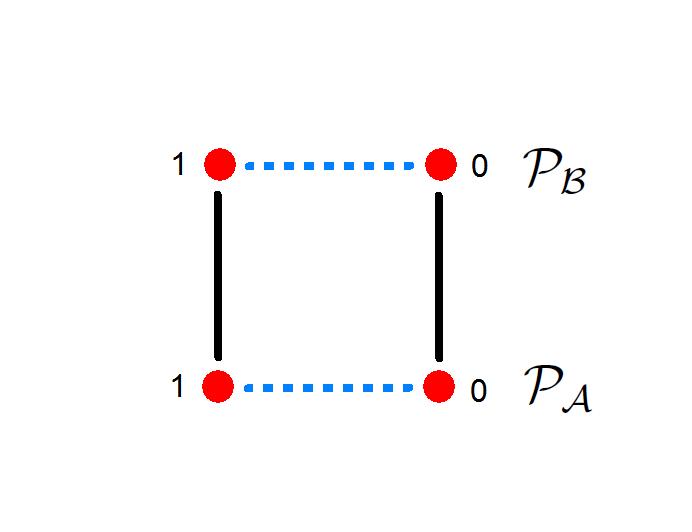}}
\subfigure[\label{Fig6b}]{
\includegraphics[width=3.5cm]{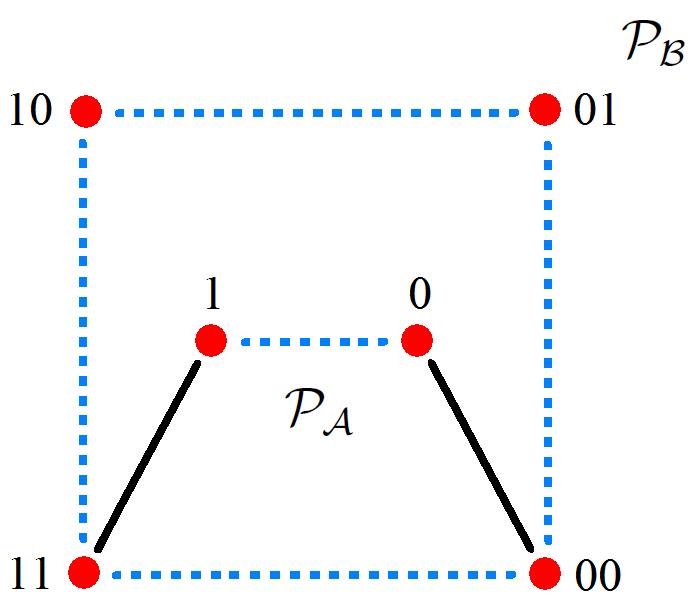}}
\subfigure[\label{Fig6c}]{
\includegraphics[width=3.5cm]{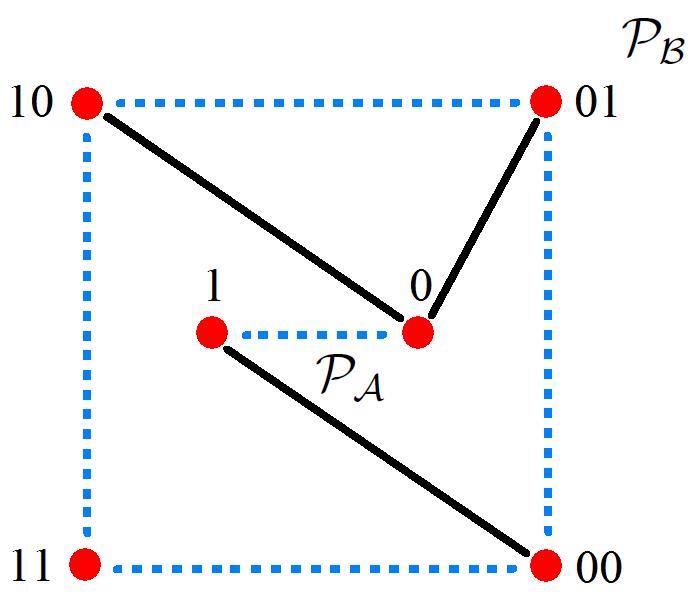}}
\subfigure[\label{Fig6d}]{
\includegraphics[width=3.5cm]{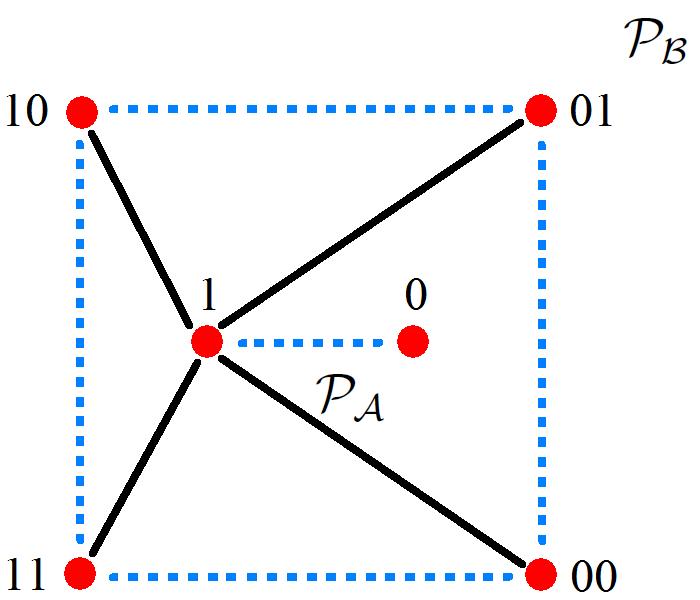}}
\caption{Graph representation of certain states of systems composed of (a) two, and (b)-(d) three qubits, where (a) is a Bell state, (b) is the GHZ state, (c) is the W state and (d) is the separable state given in Eq.(\ref{sep}). Auxiliary dashed lines do not belong to the graph as they represent polygons ${\cal P}_A$ and ${\cal P}_B$. States (a) and (b) represent $1$--uniform states while states (c) and (d) are not, as they do not satisfy the diagonality (A')  and uniformity (B') conditions.}
\end{center}
\label{Fig6}
\end{figure}

These conditions must hold for every bipartition of the state. In Fig. \ref{Fig6} we show some graphs associated
to known pure states: the Bell state $|\Phi^{+}_{2}\rangle$ of (\ref{Bell2}),
the GHZ state (\ref{GHZ3}) and the $3$--qubit $W$ state (\ref{W3}). Given that edges define the degree of entanglement of a state one could speculate
 that the entanglement increases with the number of edges.
However, it is not so, as  some separable states correspond to the graphs with several edges.
For example, the graph shown in Fig. 4d 
 represents the following separable state:
\begin{eqnarray}\label{sep}
|\Phi_{sep}\rangle&=&|100\rangle+|101\rangle+|110\rangle+|111\rangle\nonumber\\
&=&|1\rangle\left(|00\rangle+|01\rangle+|10\rangle+|11\rangle\right),\nonumber\\
\end{eqnarray}
where the first qubit is not entangled to the rest. This state reflects the property 3. mentioned above. By construction, every graph uniquely  determines a pure state. However, the graph representation depends on the bipartition considered, in general. Therefore, for constructing $k$--uniform states from graphs we should follow the above rules A' and B' for every bipartition, which is a complicated task for a high number of qudits. In fact, there are $\binom{N}{k}$ graphs associated to every $k$--uniform state of $N$ qudits. Interestingly, $k$--uniform states having entries from the set $\{0,1\}$ have all its graphs \emph{isomorphic}. We recall that two graphs $G$ and $H$ are isomorphic if there exist an isomorphism $f$ such that every pair of vertices $u,v\in G$ are adjacent if and only if $f(u),f(v)\in H$ are adjacent. This property holds because the subset of $k$--uniform states considered is one-to-one connected with OA. In particular, if a $k$--uniform state comes from an OA of index unity then all its
graphs are identical. This property has a straightforward
proof from definition of orthogonal arrays.

A graph $G$ associated to a $k$--uniform state has a simple representation as a function of its adjacency matrix $M^G$. Here, $M^G_{ij}=1$ if the vertex $V_i\in\mathcal{P}_A$ is connected to the vertex $V_j\in\mathcal{P}_B$ and it is zero otherwise. As we have shown in Fig.\ref{Fig6}, the indices $i$ and $j$ are labeled by \emph{d}--inary numbers. Therefore, the state associated to a matrix $M^G$ is given by:
\begin{equation}
|\Phi\rangle=\sum_{i,j}M^G_{ij}|ij\rangle,
\end{equation}
where $i$ and $j$ take all the values such that $V_i\in\mathcal{P}_A$ and $V_j\in\mathcal{P}_B$.

 A final comment deserves our attention here. The graph representation
used here allows us to distinguish completely disentangled states of $d$--level systems (see Fig. 4d) 
 form genuine multipartite entangled states. Furthermore, the structure of the graphs associated with pure quantum states suggests to divide them into following classes:
\begin{enumerate}
\item[\emph{(i)}] fully separable
\item[\emph{(ii)}] Partially entangled
\item[\emph{(iii)}] Genuinely entangled
\item[\emph{(iv)}] $k$--uniform.
\end{enumerate}
In the last case, the strength $k$ is also encoded in the graph.
We encourage to the reader to verify that
the rules \emph{(A')} and \emph{(B')}
 are satisfied for any GHZ state of $N$ qudits.

\subsection{Quantum gate allowing one to generate a \textbf{\emph{k}}--uniform state}
We describe here how to generate $k$--uniform states by applying quantum gates to a
selected initial \emph{blank} state:
\begin{equation}
|\eta_0\rangle=|0\rangle^{\otimes N}.
\end{equation}

A general recipe to generate a $k$--uniform states reads:
\begin{enumerate}
\item Apply the local operation $H^{\otimes(k)}\otimes\mathbb{I}^{\otimes(N-k)}$,
where $H$ is the single--qubit Hadamard gate and expand the terms involving the first $k$ qudits.
That is create the state
\begin{equation}
|\eta_1\rangle=\left(\sum_{s_1,\dots,s_k=0}^d|s_1,\dots,s_k\rangle\right)|0\rangle^{\otimes (N-k)},
\end{equation}
with $r=d^k$ terms.

\item Apply a unitary transformation satisfying the following restriction:
\begin{equation}\label{U}
U\left(|i\rangle\otimes|0\rangle^{\otimes (N-k)}\right)=|i\rangle\otimes\sum_{j=0}^{\lambda-1}|f_j(i)\rangle,
\end{equation}
\end{enumerate}
for every $i$ such that $V_i\in\mathcal{P}_A$.
A pair $\{i,f_j(i)\}$ denotes two connected vertices $V_{i}\in\mathcal{P}_A$ and $V_{f_j(i)}\in\mathcal{P}_B$,
 where $j=0,\dots,\lambda-1$. Here we require $\lambda$ functions $f$ because every vertex of the polygon
$\mathcal{P}_A$ has exactly $\lambda$ connections to a vertex of $\mathcal{P}_B$, where $\lambda$ is the index of the OA associated to the $k$--uniform state. Note that $U$ is well--defined because the inner product between vectors in the domain is
 preserved in the image. Additionally, this operator clones partial information stored in orthogonal states,
so it does not contain a universal cloning machine for any pair of qudits.
In general, $U$ can be always decomposed as simple non-local unitaries. For instance, to
prepare the GHZ state of $N$ qubits we use a generalized control NOT gate,
with a single control qubit and $N-1$ target qubits. That is,
\begin{equation}
U = {\mathbbm 1}_2^{\otimes N-1} \oplus \sigma_x^{\otimes N-1}.
\end{equation}

Finally, the step $1.$ of our recipe has only an illustrative meaning, in order to have a smooth connection with our scheme. Thus, it can be removed in practice because it only involves local operations.

\subsection{Mutually unbiased bases and \textbf{\emph{k}}--uniform states}
The construction of mutually unbiased bases (MUB) has become an intriguing problem in pure mathematics and it has
some important applications to quantum tomography. MUB are naturally connected with other interesting hard problems:
affine planes, mutually orthogonal Latin squares and complex Hadamard matrices \cite{DEBZ10}.
Here, our intention is to connect MUB with $k$--uniform states.
Specifically, we will relate \emph{single} multipartite quantum states with maximal sets of MUB.
Two orthonormal bases $\{|\varphi_s\rangle\}$ and $\{|\phi_t\rangle\}$ defined on a $d$-dimensional Hilbert space are
{\sl mutually unbiased}
 if
\begin{equation}
|\langle\varphi_s|\phi_t\rangle|^2=\frac{1}{d},
\end{equation}
for every $s,t=0,\dots,d-1$. Maximal set of $(d+1)$ MUB exists in every prime power dimension \cite{Wootters} and
in spite of many efforts its existence remains open for every composite dimension.
In order to connect MUB with $k$--uniform states we consider a reduced version of
Theorem 8.43 (see p.. 192, \cite{Hedayat}):

\begin{teo}[Hedayat]\label{TeoHed}
An OA($d^2,d+1,d,2$) exists if and only if a projective plane of order $d$ exists.
\end{teo}
On one hand, the orthogonal arrays considered in this theorem have index unity (leading us to $k$-uniform states).
On the other hand, it is conjectured that the existence of a projective plane of order $d$ determines a maximal set of $(d+1)$ MUBs in dimension $d$ \cite{Saniga}.
A confirmation of this result together with Theorems \ref{OAkMM} and \ref{TeoHed} would imply the following interesting result: If a $2$--uniform state of $d+1$ qudits having $d^2$ terms exists then a maximal set of $d+1$ mutually unbiased bases exist in dimension $d$. Note that $2$--uniform states of $7$ subsystems of 6 levels each composed of $36$ terms do not exist because of projective planes of order 6 do not exist. A similar conclusion can be done from the non-existence of a projective plane of order 10.

In the subsequent section we present illustrative examples of our
construction of $k$--uniform states.

\section{Examples of $k$--uniform multipartite states}
To show our construction in action we present in this section some simple examples of $k$--uniform states connected to orthogonal arrays. Working with quantum states we use the standard notation, but for brevity we work with not normalized pure
states. Let us start with the simplest orthogonal array
\begin{equation}\label{OA2Q}
OA(2,2,2,1)=\begin{array}{cc}
    0 & 1\\
    1 & 0
  \end{array}.
\end{equation}
In a combinatorial context we deal with $2\times2$ orthogonal array of $d=2$ levels and strength $k=1$.
As defined in Section III, the strength denotes the number of columns that should be considered in order
 to have $\lambda=1$ times repeated a sequence of $k=1$ symbols along the rows.
In the physical context, OA(2,2,2,1) represents a quantum superposition of qubits containing 2 terms,
each representing a separable pure state of two qubits, $d=2$, which is a $1$--uniform state.
The strength $k$ of the array determines the class of the  $k$--uniform state.
Thus, the state connected with the array (\ref{OA2Q}) is the maximally entangled Bell state,
\begin{equation}
|\Psi^+_2\rangle=|01\rangle+|10\rangle.
\end{equation}
In this section we use orthogonal arrays listed in the resource of Sloane \cite{Sloane}.
In this catalog, we find the following orthogonal array:
\begin{equation}\label{OA3Q}
OA(4,3,2,2)=\begin{array}{ccc}
0&0&0\\
0&1&1\\
1&0&1\\
1&1&0
\end{array}.
\end{equation}
It cannot be used to construct a $2$--uniform state of $N=3$ qubits since inequality $k\leq N/2$ does not hold (see Theorem \ref{OAkMM}).
However, an OA($r,N,d,k$) is always an OA($r,N,d,k-1$), so we can consider OA(4,3,2,2) as OA(4,3,2,1), which is an IrOA.
Then inequality $k\leq N/2$ holds and the following $1$--uniform, balanced
 state of three qubits arises:
\begin{equation}
\label{Phi3}
|\Phi_3\rangle=|000\rangle+|011\rangle+|101\rangle+|110\rangle.
\end{equation}
Acting on this state with a tensor product of Hadamard matrices $H_2\otimes H_2 \otimes H_2$
one can verify that it is locally equivalent to the standard GHZ state (\ref{GHZ3}). Interestingly, this three--qubit GHZ state defines the OA
\begin{equation}\label{OA3QGHZ}
OA(2,3,2,1)=\begin{array}{ccc}
0&0&0\\
1&1&1
\end{array}.
\end{equation}
Orthogonal arrays (\ref{OA3Q}) and (\ref{OA3QGHZ}) do not have anything in common
in the theory orthogonal arrays. Thus a natural question arises:
which OA are related by the fact that the corresponding quantum states are
equivalent with respect to local unitary transformations?

Let us now consider the following array,
\begin{equation}
\label{OA4Q}
\mbox{OA}(8,4,2,3)=\begin{array}{cccc}
0&0&0&0\\
0&0&1&1\\
0&1&0&1\\
0&1&1&0\\
1&0&0&1\\
1&0&1&0\\
1&1&0&0\\
1&1&1&1\\
\end{array},
\end{equation}
which is also an OA(8,4,2,1).  This array leads to the following $1$--uniform state of 4 qubits,
\begin{eqnarray}
\label{Phi4}
|\Phi_4\rangle&=&|0000\rangle+|0011\rangle+|0101\rangle+|0110\rangle+\nonumber\\
&&|1001\rangle+|1010\rangle+|1100\rangle+|1111\rangle ,\nonumber\\
\end{eqnarray}
well known in the literature \cite{DVC00}.
 It is worth to add that a generalization of the above state
 with complex coefficients expressed by the third root of unity
is known in the literature as $L$--state, for which certain measures
of quantum entanglement achieve its maximum \cite{Gour}. Recently it was
proved \cite{CD13} that this state yields the maximum
of the absolute value of the hyperdeterminant for 4 qubits.

Observe that existence of an OA(4,4,2,2) would immediately lead us to a $2$--uniform state of 4 qubits.
 However, the Rao bound tells us that an OA($r$,4,2,2) must satisfy $r\geq5$.
 Note that for arrays with $\lambda \ge 2$, the diagonality condition $B$) is not always satisfied,
so not every such array allows us to construct the corresponding $k$--uniform state.
 For example, from the array with the index $\lambda=2$,
\begin{equation}
\label{OA5Q}
\mbox{OA}(8,5,2,2)=\begin{array}{ccccc}
0&0&0&0&0\\
1&0&0&1&1\\
0&1&0&1&0\\
0&0&1&0&1\\
1&1&0&0&1\\
1&0&1&1&0\\
0&1&1&1&1\\
1&1&1&0&0
\end{array},
\end{equation}
we obtain the following $1$--uniform state of 5 qubits:
\begin{eqnarray}\label{Phi5}
|\Psi_5\rangle&=&|00000\rangle+|10011\rangle+|01010\rangle+\nonumber\\&&|00101\rangle+|11001\rangle+|10110\rangle+\nonumber\\&&|01111\rangle+|11100\rangle.
\end{eqnarray}
It is easy to verify that this state is not $2$--uniform.
The only non--maximally mixed reductions to two qubits are $\rho_{24}$ and $\rho_{35}$.
 Curiously, we could not find a $2$--uniform state of 5 qubits having $\{0,1\}$ entries and
it is likely,  they do not exist.
In Appendix \ref{AppendixC} we derive a $2$--uniform state of 5 qubits having $\{0,\pm1\}$ entries.
Remarkably, for any number of qubits $N>5$ it is possible to find a $2$--uniform state having $\{0,1\}$ entries, as
 we will show in Theorem \ref{Had2MM} (see also Appendix \ref{AppendixA}).

\begin{figure}[!h]
\centering 
{\includegraphics[width=8cm]{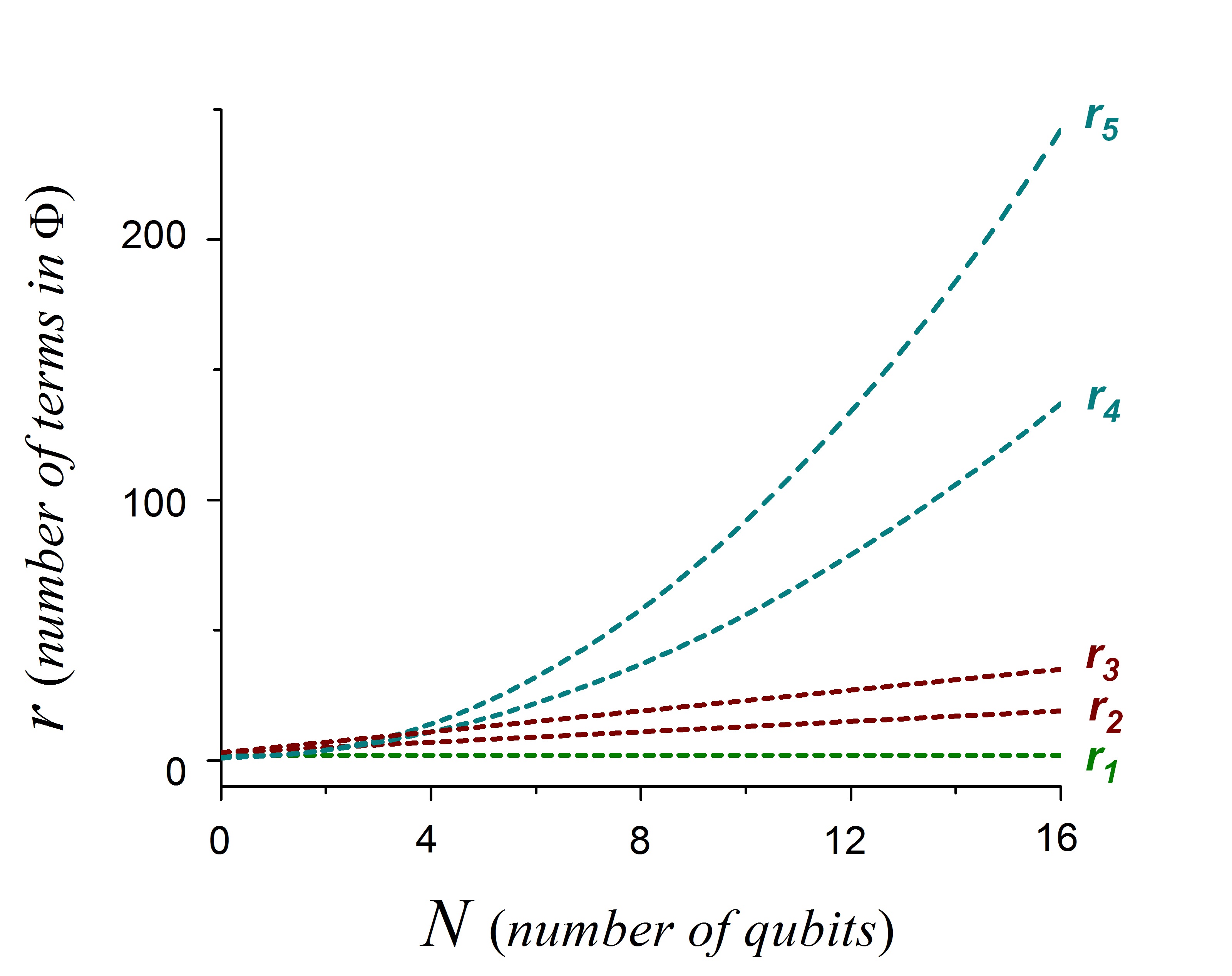}} 
\caption{Lower bound for the number of terms $r_k$ of a
$k$--uniform state of $N$ qubits imposed by the Rao bounds (\ref{raobounds}):
$r_{1}=2$, $r_{2}=N+1$, $r_{3}=2N$, $r_{4}=N^2/2+N/2+1$ and $r_{5}=N^2-N+2$.}
\label{Fig3}
\end{figure}

In the case of a higher number of qubits the number of terms $r$ in a $k$--uniform state becomes large
according to the Rao bound (\ref{raobounds}). In Fig. \ref{Fig3} we show how the minimal number of terms $r$ required to produce a $k$--uniform pure state of $N$ qubits depends on the number $N$.
The following remarks arise:
\begin{enumerate}
\item For $1$--uniform states we have $r\geq2$, as there exist a family of generalized GHZ states of $N$ qubits,
which consist of two terms. From a combinatorial point of view, the arrays
corresponding to multi--qubit GHZ states were discussed by Hedayat \cite{Hedayat} (Example 1.4, p. 3)
as a trivial way to construct OA(r,N,2,1).

\item Rao bounds tells us that $1$--uniform states is the only kind of $k$--uniform states allowing a constant number of terms, $r_1=2$.

\item The minimum number of terms $r_2$ and $r_3$ required to have a $2$--uniform or $3$--uniform state respectively,
is a growing \emph{linear function} of the number of qubits $N$, with slopes $1$ and $2$.

\item The minimum number of terms $r$ required to have a $k$--uniform state for $k>3$ is a growing
\emph{non-linear function} of the number of qubits $N$. For example, in the case of $k=4$
 we have $r_4\geq1+N/2+N^2/2$ and $r_5 \geq 2-N+N^2$.
\end{enumerate}

\section{Construction of $\mathbf{2}$--uniform states for every $\mathbf{N>4}$}\label{2MM}

Generalized GHZ states provide a simple example of $1$--uniform states for an arbitrary number of $N$ qubits. Construction of $2$--uniform states usually requires computing
simulations and numerical approximations --
see \cite{Borras} and references therein. In fact, $2$--uniform states are explicitly known for 5 and 6 qubits only \cite{Arnaud}.

In this section we solve the problem of constructing a kind of $2$--uniform states for an arbitrary number of $N$ qubits. As it is often the case \cite{We01}, the theory of quantum information
can benefit from combinatorics and Hadamard matrices.
 A Hadamard matrix is up to a prefactor an orthogonal matrix with entries $\pm1$.
 Some orthogonal arrays of strength 2 are connected with
the famous Hadamard conjecture
-- see Theorem 7.5, p. 148 \cite{Hedayat}:

\begin{teo}(Hedayat)\label{teoHad}
Orthogonal arrays OA($4\lambda,4\lambda-1,2,2$) exists if and only if there exists a Hadamard matrix of order $4\lambda$.
\end{teo}
This theorem combined with a partial classification of Hadamard matrices allows us to find a kind of $2$--uniform states of an arbitrarily large number of qubits.
Consider a Hadamard matrix of order $\kappa$ in normalized form,
so that the entries appearing in the first row and the first column are equal to unity.

Making use of Theorem \ref{teoHad} we realize that Hadamard matrices could be useful to construct $2$--uniform states
of $\kappa-1$ (or less) qubits. That is, we will be able to find an OA which satisfies the diagonality property
 \emph{(B)}. In what follows we derive, for which values of $N$ we can construct a $2$--uniform state from a given
Hadamard matrix of size $\kappa$. On one hand, every entry of the first row of the OA contains ones.
Thus, we must avoid to have another row containing $N-2$ ones since, according to \emph{(B)}
we need to have different rows for every subset of $N-2$ columns.
Given that we work with normalized Hadamard matrices, zeros and ones appear \emph{at most} $\kappa/2-1$ times in every row,
excluding the first one.
Therefore, if the number of qubits to be reduced ($N-2$) is greater than $\kappa/2-1$ it is not possible to obtain
 a second row of ones.
Additionally, if the first row is not repeated then the remaining rows are not repeated either,
 as in OA every repeated pair of rows is repeated the same number of times.
Thus, the $N-2$ columns to be reduced are different iff $\kappa/2-1<N-2$ and thus property (B) holds. Given that $N\leq\kappa-1$ we
arrive at the following statement.
\begin{teo}\label{Had2MM}
A Hadamard matrix of order $\kappa$ allows one to find $2$--uniform states of $N$ qubits having entries from the set $\{0,1\}$ iff
\begin{equation}\label{cotakappa0}
\kappa/2+2\leq N\leq\kappa-1.
\end{equation}
\end{teo}
In Appendix \ref{AppendixA} we exemplify this theorem by constructing $2$--uniform states for 6 to 15 qubits.
\begin{figure}[!h]
\centering 
{\includegraphics[width=8cm]{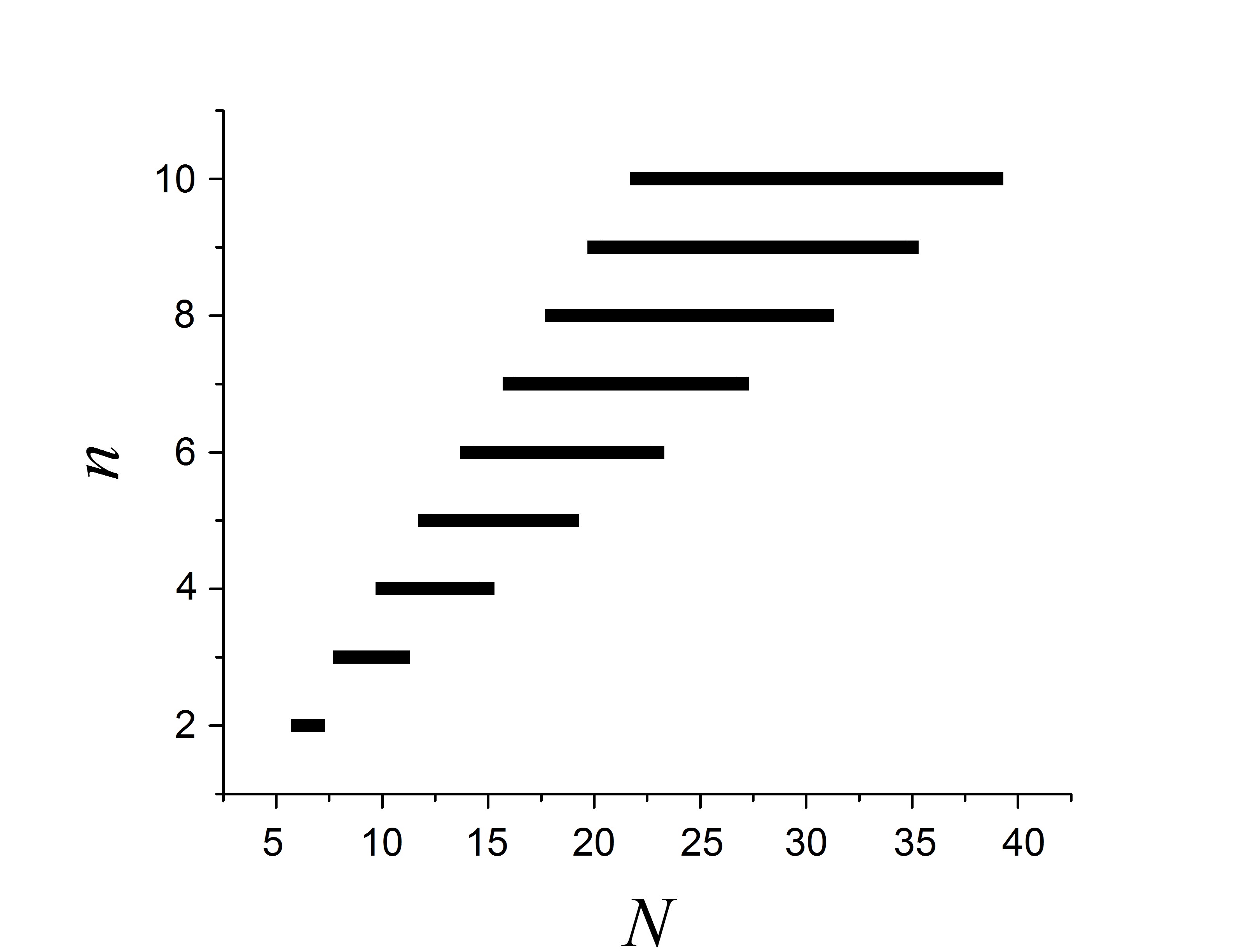}} 
\caption{Overlap of the bounds $3\times2^{n-2}+2 \ \leq \  N\ \leq \ 3\times2^{n-1}$\\ (See Eq.(\ref{cotakappa})).}
\label{Fig4}
\end{figure}

Combinatorial Theorem \ref{teoHad} can be thus adopted to the theory of multipartite entanglement.
Together with the above explanations it leads us to the following result:
\begin{teo}\label{MMHad}
The problem of classifying $2$--uniform states for qubits contains the Hadamard conjecture.
\end{teo}
Precisely, the 2--uniform states of $N=4\lambda-1$ qubits connected with the Hadamard conjecture are those having $r=4\lambda$ terms ($\lambda\in\mathbb{N}$). The Hadamard conjecture states that a $\kappa\times\kappa$ Hadamard matrix exists for $\kappa=2$ and for every $\kappa=4n$, $n\in\mathbb{N}$. It is open since 1893 and it represents one of the most important problems in combinatorics. Not knowing, whether the Hadamard conjecture holds, we show that Theorem \ref{Had2MM}
allows us to construct another kind of $2$--uniform states for an arbitrary number of $N> 4$ qubits.

 First, we note that $\kappa=2^n$ ($n\in\mathbb{N}$) is not a  solution
for every $N$. That is, the intervals given in Eq.(\ref{cotakappa0})
 have gaps for some values of $N$.
Specifically, the gaps are given for $N=2^n$ and $N=2^n+1$ (for every $n>2$).
In order to cover these gaps it is enough to consider $\kappa=12\times2^{n-3}$, where the new interval is given by
\begin{equation}
\label{cotakappa}
3\times2^{n-2}+2 \ \leq \  N\ \leq \ 3\times2^{n-1}.
\end{equation}

Fig. \ref{Fig4} shows that many pairs $(n,N)$ satisfying Eq.(\ref{cotakappa}) exist for every $N>5$. As Hadamard matrices of the size  $\kappa=12\times2^{n-3}$ can be written down explicitly as a tensor product
$(H_{12} \otimes H_2^{\otimes n-3})$
we are able to construct 2-uniform states of $N$ qubits for $N > 5$. Thus, the following statement holds:
\begin{teo}\label{teo2MMQ}
For every $N>5$ there exists $2$--uniform quantum states of $N$ qubits and they can be constructed from known Hadamard matrices. Moreover, they have entries from the set $\{0,1\}$.
\end{teo}
The proof of the above theorem is constructive:
\begin{itemize}
\item [\emph{(i)}] Consider a $\kappa\times\kappa$  Hadamard matrix $H$ in the normalized form, so that
its first column and first row consist of $+1$ only.
 If $N\neq2^n$ and $N\neq2^n+1$ for every $n>2$ then consider $\kappa=2^\nu$,
where $\nu$ is chosen such that $\kappa-N$ is positive and, for simplicity, is as small as possible.
Else, consider $\kappa=12\times2^{\nu-3}$ in the same way.

\item [\emph{(ii)}] Discard the first column of $H$, keep any subset of $N$ columns and discard the rest
       to obtain a rectangular matrix with $\kappa$ rows and $N$ columns.

\item [\emph{(iii)}] Replace all elements equal to $-1s$ by $0$. This lead us to an OA according to Theorem \ref{teoHad}
    and property \emph{(v)} from section \ref{OAandKMM}.
\item [\emph{(iv)}] Every row of the OA obtained should be put in kets, summed and normalized
to produce the desired $2$--uniform state of $N$ qubits, which completes the construction. \qed
\end{itemize}
\medskip
Note that for $\lambda=1$ in Theorem \ref{teoHad} we have an OA(4,3,2,2) which cannot be used to construct a $2$--uniform state of 3 qubits
as condition $k\leq N/2$ is not satisfied. Instead, we have shown in Eq.(\ref{OA3Q})
that this array allows one to build a $1$--uniform state of 3 qubits.

As final comments we realize that the complete classification of $2$--uniform states is currently out of reach, as
it includes the Hadamard conjecture. Moreover, already for $\kappa=32$
 there are more than 13 millions of non-equivalent Hadamard matrices \cite{Kharaghani},
so the number of ways to generate $2$--uniform states for 31 qubits is huge.
We do not know how many of these states are locally equivalent.

Another observation is worth to be made:
We realize that a  $k$--uniform state constructed
from IrOA according to the scheme shown in Fig. \ref{Fig2}
allows us to generate entire classes of $k$--uniform states. For instance, taking
\begin{equation}
\label{Phi3b}
|\Phi_3\rangle=|000\rangle+|011\rangle+|101\rangle+|110\rangle,
\end{equation}
we generate the three-parameter class of states
\begin{eqnarray}
\label{Phi3c}
|\Phi_3\rangle(\alpha_1,\alpha_2,\alpha_3)&=&|000\rangle+e^{i\alpha_1}|011\rangle+\nonumber\\&&e^{i\alpha_2}|101\rangle+e^{i\alpha_3}|110\rangle,\nonumber\\
\end{eqnarray}
as $\alpha_1,\alpha_2,\alpha_3\in[0,2\pi)$.
In general, from any $k$--uniform state of $N$ qudits generated by an OA of index $\lambda=1$, i.e.,
\begin{eqnarray}
|\Phi\rangle&=&|s^1_1,s^1_2,\dots,s^1_N\rangle+|s^2_1,s^2_2,\dots,s^2_N\rangle+\nonumber\\&&\dots+|s^r_1,s^r_2,\dots,s^r_N\rangle,
\end{eqnarray}
we can generate the entire $d^N-1$ dimensional orbit of $k$--uniform quantum states,
\begin{eqnarray}
\label{orbit46}
|\Phi\rangle&=&|s^1_1,s^1_2,\dots,s^1_N\rangle+e^{i\alpha_1}|s^2_1,s^2_2,\dots,s^2_N\rangle+\nonumber\\&&
\dots+e^{i\alpha_{d^N-1}}|s^r_1,s^r_2,\dots,s^r_N\rangle .
\end{eqnarray}
Therefore, our approach is not only useful to generate $k$--uniform state but also to generate
entire orbits of maximally entangled states. It is easy to see that an orbit of $k$--uniform
states can be considered as a starting point for a search for
a single $(k+1)$--uniform state.

\section{Concluding remarks}
We presented a combinatorial tool to systematically generate genuine multipartite entangled pure states of $N$ qudits:
the orthogonal arrays. Our construction of $k$--uniform states
works if an orthogonal array of strength $k$ is
irredundant (i.e., after removing from the array any $k$ columns
all remaining rows are distinct). Note that this class of orthogonal
arrays differs from the super-simple orthogonal arrays,
recently discussed in \cite{STY12}.

Our approach allows us to find new entangled states of several
qudits and to establish their properties.
 In Section \ref{OAandKMM} we derived two conditions sufficient to construct
 $k$--uniform states and realized that
  the first one, uniformity condition (A), is fulfilled by any orthogonal array.
  Having at hand any orthogonal array one needs to verify that
it is irredundant, so the diagonality condition (B) is satisfied.
  In practice, it is easy to find examples of arrays
for which both conditions  are satisfied,
 so they lead to $k$--uniform states.

Results obtained in this paper include:
\begin{enumerate}
\item Every orthogonal array of index unity OA($d^k,N,d,k$) allows us to generate a $k$--uniform state of
 $N$ qudits of $d$ levels if and only if $k\leq N/2$ -- see Theorem \ref{OAkMM}.
In Appendix \ref{AppendixB} we present
 several examples of $k$--uniform states for subsystems with $d>2$ levels each.

\item We demonstrated that $2$--uniform states of $N$ qubits exist for every $N>4$,  see Section \ref{2MM}.
    An explicit construction of these states is presented in
  Theorem \ref{teo2MMQ} which
    involves known Hadamard matrices.
 Such states are listed in Appendix \ref{AppendixA} for $N=6,\dots, 15$.

\item We revisited the connection between $k$--uniform states and quantum error correction codes. In particular, we find that maximal distance separable \emph{classical} codes are equivalent to some $k$--uniform states (see Prop. \ref{MDSCECC}) and also established connections between classical and quantum codes -- see Prop. \ref{CQECC1}.

\item A certain graph representation of $k$--uniform states is proposed.
 It arises from orthogonal arrays and it allows us to identify
the $k$--uniform states -- see Section \ref{Graph}.
Additionally, the key conditions of uniformity (A) and diagonality (B) stated on pag. \ref{AB} receive a simple
 graphical interpretation.

\item The existence of a \emph{single} $2$--uniform state of $d+1$ qudits is connected
 with the existence of a maximal set of $d+1$ mutually unbiased bases (MUBs)
in prime power dimensions (See Theorem \ref{TeoHed}). If Saniga's conjecture \cite{Saniga} is true then our connection is valid for every dimension $d$.

\item We proved that the existence of a particular kind of $2$--uniform states, those considering $N=\kappa-1$ qubits and having $\kappa$ terms are equivalent to construct $\kappa\times\kappa$ Hadamard matrices. Consequently, for $\kappa\neq 4n$ they do not exist and for $\kappa=4n$ this is equivalent to the Hadamard conjecture, for every $n\in\mathbb{N}$ (see Theorem \ref{MMHad}).
This suggests that the complete classification of $2$--uniform states of $N$--qubit states could be temporarily out of reach.

\item An entire orbit of maximally entangled states can be constructed from every $k$--uniform state generated from an OA
-- see Section \ref{2MM}.
\end{enumerate}

\medskip

\begin{figure}[!h]
\centering 
{\includegraphics[width=8cm]{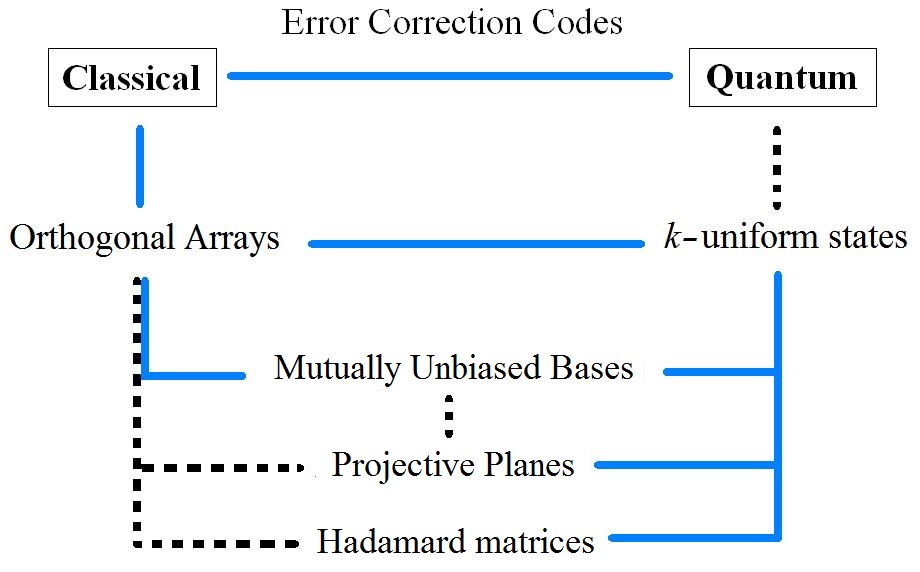}} 
\caption{Relationship between $k$--uniform states, QECC, CECC and other relevant mathematical notions.
 Dashed (black) lines represent known connections, while solid (blue) lines denote the relations discussed in this paper.}
\label{Fig7}
\end{figure}

Additionally, in Appendix \ref{AppendixC} we explain how to construct $k$--uniform states from orthogonal arrays when the diagonality assumption \emph{(B)} is not satisfied.
 In Fig.\ref{Fig7} we show the most important connections made in this work.\vspace{0.5cm}

Let us conclude this work by listing some open issues:
\begin{enumerate}
  \item[A.] Check if every $k$--uniform state is equivalent under
    SLOCC to a state generated from an orthogonal array.
         This property holds for systems containing $N=2,3,4$ qubits.

  \item[B.] Find for what $N$ there are $3$--uniform states of $N$ qubits and $2$--uniform states of $N$ qutrits. Also, provide an explicit construction of $3$--uniform states of qudits for high values of $N$.

  \item[C.] Solve the existence problem of $3$--uniform states of 7  and $8$ qubits.

  \item[D.] Find how the maximal value $k_{max}$, for which  $k_{max}$--uniform states of $N$-qubit exist, depends on $N$.
        Analyze the dependence $k_{max}(N)$ for qutrits and higher,
 $d$-dimensional systems.

  \item[E.] Investigate existence of the approximate,
 ($\epsilon, k$)--uniform states,
     for which all reductions are maximally mixed up to $\epsilon$ corrections.
    Analyze generic random pure states of $N$ qudits,
     distributed with respect to a Haar measure \cite{ZS01},
    to find, for which values of both parameters they are ($\epsilon, k$)--uniform.

  \item[F.] Extend the  method developed here to heterogeneous systems
     composed of subsystems with different number of levels,
     for instance the qubit--qutrit systems.

  \item[G.] Extend Theorem \ref{teo2MMQ} to the case of $2$--uniform states
  of qudits by using Butson--type complex Hadamard matrices \cite{TZ06}.

  \item[H.] Find what orthogonal arrays
 are {\sl locally equivalent}, \
$\mbox{OA($r,N,d,k$)} \sim_{\rm loc} \mbox{OA($r',N,d,k'$)}$,
in a sense that they lead to locally equivalent quantum states.
Note that the number $r$ of the terms in each state nor the strength $k$
 needs not to be preserved by the local equivalence relation.

\end{enumerate}

\section{Acknowledgments}
It is a pleasure to thank W. Kuhfeld for stressing the usefulness of the connection between
Hadamard matrices and orthogonal arrays, A.~Osterloh for several fruitful discussions and constructive remarks, G.~Sherwood for noting
 interesting properties on orthogonal arrays theory and to
L.~Arnaud, J.~Bielawski, D.~Braun, C.~Colbourn,  R.~Craigen, M.~Grassl, S.~Hedayat, S.~Pascazio,
D.~McNulty, J.~Stufken and P.~Zvengrowski
for very useful comments. Finally, we thank the useful comments of the referee. This work was supported by Grants FONDECyT N$^{\text{\underline{o}}}$ 3120066,
 MSI P010-30F, PIA-CONICYT PFB0824 (Chile)
and by the Maestro Grant DEC-2011/02/A/ST2/00305 financed by the Polish National Science Centre.

\appendix

\section{Explicit construction of $2$--uniform states}
\label{AppendixA}
In this section, we use known Hadamard matrices to exemplify the
construction of $2$--uniform states defined
 in Theorem \ref{teo2MMQ}. Let us consider the $8\times8$
Hadamard matrix $H_8$ and its associated
orthogonal array obtained from the last seven columns of the matrix,
\begin{eqnarray}\label{H8}
H_8=\left(
  \begin{array}{cccccccc}
1& 1& 1& 1& 1& 1& 1& 1\\
1& -1& 1& -1& 1& -1& 1& -1\\
1& 1& -1& -1& 1& 1& -1& -1\\
1& -1& -1& 1& 1& -1& -1& 1\\
1& 1& 1&1& -1& -1& -1& -1\\
1& -1& 1& -1& -1& 1& -1& 1\\
1&1& -1& -1& -1& -1& 1& 1\\
1& -1& -1& 1& -1& 1& 1& -1
  \end{array}
\right)\nonumber\\
\end{eqnarray}
\begin{eqnarray}
\rightarrow
OA=\begin{array}{ccccccc}
1& 1& 1& 1& 1& 1& 1\\
0& 1& 0& 1& 0& 1& 0\\
1& 0& 0& 1& 1& 0& 0\\
0& 0& 1& 1& 0& 0& 1\\
1& 1& 1& 0& 0& 0& 0\\
0& 1& 0& 0& 1& 0& 1\\
1& 0& 0& 0& 0& 1& 1\\
0& 0& 1& 0& 1& 1& 0
\end{array}.
\end{eqnarray}
Applying Theorem \ref{teo2MMQ} we immediately get the following $2$--uniform state of 7 qubits:
\begin{eqnarray}
|\Phi_7\rangle&=& |1111111\rangle+|0101010\rangle+|1001100\rangle+\nonumber\\
&&|0011001\rangle+|1110000\rangle+|0100101\rangle+\nonumber\\
&&|1000011\rangle+|0010110\rangle.
\end{eqnarray}
This state can be called a {\sl simplex state}, since
its eight terms, interpreted as coordinates of points in ${\mathbbm R}^7$
form a regular $7$--simplex after changing 0s by -1s \cite{AZL03}. Notice, that the scalar product
between any two such vectors is constant, so the collection of these
vectors forms equiangular lines. The same property holds also
for the states $|\Phi_{2^m-1}\rangle$, e.g. the state $|\Phi_{15}\rangle$
given in (\ref{phii15}).

Removing the first column of the OA given in Eq.(\ref{H8}) we have the following $2$--uniform state of 6 qubits:
\begin{eqnarray}
|\Phi_6\rangle&=& |111111\rangle+|101010\rangle+|001100\rangle+\nonumber\\
&&|011001\rangle+|110000\rangle+|100101\rangle+\nonumber\\
&&|000011\rangle+|010110\rangle.
\end{eqnarray}
Note that we attained the lower bound given in Eq.(\ref{cotakappa}) for the $2$--uniform states that can be directly obtained from $H_8$. However, a $2$--uniform state of 5 qubits can be obtained from $H_8$ in a different way (see Appendix C). In the following examples we systematically eliminate the first (left) qubit. From $H_{12}$ \cite{sloanehad} and Theorem \ref{teo2MMQ} we find the following $2$--uniform states:
\begin{widetext}
\begin{eqnarray}
|\Phi_{11}\rangle&=&|00000000000\rangle+|10100011101\rangle+|11010001110\rangle+|01101000111\rangle+\nonumber\\
&&|10110100011\rangle+|11011010001\rangle+|11101101000\rangle+|01110110100\rangle+\nonumber\\
&&|00111011010\rangle+|00011101101\rangle+|10001110110\rangle+|01000111011\rangle,
\end{eqnarray}

\begin{eqnarray}
|\Phi_{10}\rangle&=&|0000000000\rangle+|0100011101\rangle+|1010001110\rangle+|1101000111\rangle+\nonumber\\
&&|0110100011\rangle+|1011010001\rangle+|1101101000\rangle+|1110110100\rangle+\nonumber\\
&&|0111011010\rangle+|0011101101\rangle+|0001110110\rangle+|1000111011\rangle,
\end{eqnarray}

\begin{eqnarray}
|\Phi_{9}\rangle&=&|000000000\rangle+|100011101\rangle+|010001110\rangle+|101000111\rangle+\nonumber\\
&&|110100011\rangle+|011010001\rangle+|101101000\rangle+|110110100\rangle+\nonumber\\
&&|111011010\rangle+|011101101\rangle+|001110110\rangle+|000111011\rangle,
\end{eqnarray}
and
\begin{eqnarray}
|\Phi_{8}\rangle&=&|00000000\rangle+|00011101\rangle+|10001110\rangle+|01000111\rangle+\nonumber\\
&&|10100011\rangle+|11010001\rangle+|01101000\rangle+|10110100\rangle+\nonumber\\
&&|11011010\rangle+|11101101\rangle+|01110110\rangle+|00111011\rangle.
\end{eqnarray}\vspace{0.5cm}
Now, from the Hadamard matrix $H_{16}=H_2^{\otimes 4}$
we construct the following $2$--uniform states:
%
\begin{eqnarray}
\label{phii15}
|\Phi_{15}\rangle &=& |111111111111111\rangle+|010101010101010\rangle+|100110011001100\rangle+|001100110011001\rangle+\nonumber\\
&&|111000011110000\rangle+|010010110100101\rangle+|100001111000011\rangle+|001011010010110\rangle+\nonumber\\
&&|111111100000000\rangle+|010101001010101\rangle+|100110000110011\rangle+|001100101100110\rangle+\nonumber\\
&&|111000000001111\rangle+|010010101011010\rangle
+ |100001100111100\rangle+|001011001101001\rangle ,
\end{eqnarray}

\begin{eqnarray}
|\Phi_{14}\rangle&=&|11111111111111\rangle+|10101010101010\rangle+|00110011001100\rangle+|01100110011001\rangle+\nonumber\\
&&|11000011110000\rangle+|10010110100101\rangle+|00001111000011\rangle+|01011010010110\rangle+\nonumber\\
&&|11111100000000\rangle+|10101001010101\rangle+|00110000110011\rangle+|01100101100110\rangle+\nonumber\\
&&|11000000001111\rangle+|10010101011010\rangle+|00001100111100\rangle+|01011001101001\rangle ,
\end{eqnarray}
\begin{eqnarray}
|\Phi_{13}\rangle&=&|1111111111111\rangle+|0101010101010\rangle+|0110011001100\rangle+|1100110011001\rangle+\nonumber\\
&&|1000011110000\rangle+|0010110100101\rangle+|0001111000011\rangle+|1011010010110\rangle+\nonumber\\
&&|1111100000000\rangle+|0101001010101\rangle+|0110000110011\rangle+|1100101100110\rangle+\nonumber\\
&&|1000000001111\rangle+|0010101011010\rangle+|0001100111100\rangle+|1011001101001\rangle ,
\end{eqnarray}
\begin{eqnarray}
|\Phi_{12}\rangle&=&|111111111111\rangle+|101010101010\rangle+|110011001100\rangle+|100110011001\rangle+\nonumber\\
&&|000011110000\rangle+|010110100101\rangle+|001111000011\rangle+|011010010110\rangle+\nonumber\\
&&|111100000000\rangle+|101001010101\rangle+|110000110011\rangle+|100101100110\rangle+\nonumber\\
&&|000000001111\rangle+|010101011010\rangle+|001100111100\rangle+
|011001101001\rangle ,
\end{eqnarray}
\begin{eqnarray}
|\tilde{\Phi}_{11}\rangle&=&|11111111111\rangle+|01010101010\rangle+|10011001100\rangle+|00110011001\rangle+\nonumber\\
&&|00011110000\rangle+|10110100101\rangle+|01111000011\rangle+|11010010110\rangle+\nonumber\\
&&|11100000000\rangle+|01001010101\rangle+|10000110011\rangle+|00101100110\rangle+\nonumber\\
&&|00000001111\rangle+|10101011010\rangle+|01100111100\rangle+
|11001101001\rangle ,
\end{eqnarray}
\begin{eqnarray}
|\tilde{\Phi}_{10}\rangle&=&|1111111111\rangle+|1010101010\rangle+|0011001100\rangle+|0110011001\rangle+\nonumber\\
&&|0011110000\rangle+|0110100101\rangle+|1111000011\rangle+|1010010110\rangle+\nonumber\\
&&|1100000000\rangle+|1001010101\rangle+|0000110011\rangle+|0101100110\rangle+\nonumber\\
&&|0000001111\rangle+|0101011010\rangle+|1100111100\rangle+|1001101001
\rangle ,
\end{eqnarray}
\begin{eqnarray}
|\tilde{\Phi}_{9}\rangle&=&|111111111\rangle+|010101010\rangle+|011001100\rangle+|110011001\rangle+\nonumber\\
&&|011110000\rangle+|110100101\rangle+|111000011\rangle+|010010110\rangle+\nonumber\\
&&|100000000\rangle+|001010101\rangle+|000110011\rangle+|101100110\rangle+\nonumber\\
&&|000001111\rangle+|101011010\rangle+|100111100\rangle+|001101001\rangle ,
\end{eqnarray}
and
\begin{eqnarray}
|\tilde{\Phi}_{8}\rangle&=&|11111111\rangle+|10101010\rangle+|11001100\rangle+|10011001\rangle+\nonumber\\
&&|11110000\rangle+|10100101\rangle+|11000011\rangle+|10010110\rangle+\nonumber\\
&&|00000000\rangle+|01010101\rangle+|00110011\rangle+|01100110\rangle+\nonumber\\
&&|00001111\rangle+|01011010\rangle+|00111100\rangle+|01101001\rangle\nonumber.
\end{eqnarray}
\end{widetext}

In the case of $N=8,9,10,11$ we decorated the states by a symbol~ ${}\tilde{}$
~ in order to emphasize that the states obtained from the Hadamard matrix $H_{16}$ need not
to coincide with these constructed from $H_8$.
Furthermore, for $N=16$  there exists 5 different
classes of non-equivalent Hadamard matrices \cite{sloanehad},
and all of them allow us to generate $2$--uniform states for
$N\le 15$ qubits. Finally, we remark that each of the $2$--uniform states of $k$ qubits generated from our method leads
us to the orbits of $2$--uniform states with $2^k-1$ real parameters
constructed according to Eq. (\ref{orbit46}).

\section{$\mathbf{k}$--uniform states for $d$--level subsystems} \label{AppendixB}
In this section we construct $k$--uniform states from orthogonal arrays for qudits having $d>2$. Let us present some $k$--uniform states straightforwardly constructed from OA of index $\lambda=1$. The following OA have been taken from the catalog of Sloane \cite{Sloane}. From OA(9,4,3,2) we have the $2$--uniform of 4 qutrits:
\begin{eqnarray}
|\Psi_3^4\rangle&=&|0000\rangle+|0112\rangle+|0221\rangle+\nonumber\\
&&|1011\rangle+|1120\rangle+|1202\rangle+\nonumber\\
&&|2022\rangle+|2101\rangle+|2210\rangle.
\end{eqnarray}
Interestingly, a $2$--uniform  state of $4$ qutrits exists whereas there is no
$2$--uniform state of $4$ qubits. From OA(16,5,4,2) we get the $2$--uniform state of 5 ququarts,
related to the Reed--Solomon code \cite{Sloane} of length $5$,
\begin{widetext}
\begin{eqnarray}
|\Psi_4^5\rangle&=&|00000\rangle+|01111\rangle+|02222\rangle+|03333\rangle+|10123\rangle+|11032\rangle+|12301\rangle+|13210\rangle+\nonumber\\
&&|20231\rangle+|21320\rangle+|22013\rangle+|23102\rangle+|30312\rangle+|31203\rangle+|32130\rangle+|33021\rangle.\nonumber\\
\end{eqnarray}
\end{widetext}

From OA(64,6,4,3) we obtain a $3$--uniform state of $6$ ququarts:
\begin{widetext}
\begin{eqnarray}
|\Psi_4^6\rangle&=&|000000\rangle+|001111\rangle+|002222\rangle+|003333\rangle+|010123\rangle+|011032\rangle+\nonumber\\
&&|012301\rangle+|013210\rangle+|020231\rangle+|021320\rangle+|022013\rangle+|023102\rangle+\nonumber\\
&&|030312\rangle+|031203\rangle+|032130\rangle+|033021\rangle+|100132\rangle+|101023\rangle+\nonumber\\
&&|102310\rangle+|103201\rangle+|110011\rangle+|111100\rangle+|112233\rangle+|113322\rangle+\nonumber\\
&&|120303\rangle+|121212\rangle+|122121\rangle+|123030\rangle+|130220\rangle+|131331\rangle+\nonumber\\
&&|132002\rangle+|133113\rangle+|200213\rangle+|201302\rangle+|202031\rangle+|203120\rangle+\nonumber\\
&&|210330\rangle+|211221\rangle+|212112\rangle+|213003\rangle+|220022\rangle+|221133\rangle+\nonumber\\
&&|222200\rangle+|223311\rangle+|230101\rangle+|231010\rangle+|232323\rangle+|233232\rangle+\nonumber\\
&&|300321\rangle+|301230\rangle+|302103\rangle+|303012\rangle+|310202\rangle+|311313\rangle+\nonumber\\
&&|312020\rangle+|313131\rangle+|320110\rangle+|321001\rangle+|322332\rangle+|323223\rangle+\nonumber\\
&&|330033\rangle+|331122\rangle+|332211\rangle+|333300\rangle.
\end{eqnarray}
\end{widetext}

From OA(25,6,5,2) we get a $2$--uniform state consisting of six $5$--level systems
\begin{widetext}
\begin{eqnarray}
|\Psi_5^6\rangle&=&|000000\rangle+|011234\rangle+|022341\rangle+|033412\rangle+|044123\rangle+\nonumber\\
&&|101111\rangle+|112403\rangle+|124032\rangle+|130324\rangle+|143240\rangle+\nonumber\\
&&|202222\rangle+|214310\rangle+|223104\rangle+|231043\rangle+|240431\rangle+\nonumber\\
&&|303333\rangle+|310142\rangle+|321420\rangle+|334201\rangle+|342014\rangle+\nonumber\\
&&|404444\rangle+|413021\rangle+|420213\rangle+|432130\rangle+|441302\rangle.
\end{eqnarray}
\end{widetext}

Interestingly, the $1$--uniform state of 7 qubits obtained from OA(64,7,2,6)
\begin{widetext}
\begin{eqnarray}
|\Psi_2^7\rangle&=&|0000000\rangle+|1000001\rangle+|0100001\rangle+|0010001\rangle+|0001001\rangle+|0000101\rangle+\nonumber\\
&&|0000011\rangle+|1100000\rangle+|1010000\rangle+|1001000\rangle+|1000100\rangle+|1000010\rangle+\nonumber\\
&&|0110000\rangle+|0101000\rangle+|0100100\rangle+|0100010\rangle+|0011000\rangle+|0010100\rangle+\nonumber\\
&&|0010010\rangle+|0001100\rangle+|0001010\rangle+|0000110\rangle+|1110001\rangle+|1101001\rangle+\nonumber\\
&&|1100101\rangle+|1100011\rangle+|1011001\rangle+|1010101\rangle+|1010011\rangle+|1001101\rangle+\nonumber\\
&&|1001011\rangle+|1000111\rangle+|0111001\rangle+|0110101\rangle+|0110011\rangle+|0101101\rangle+\nonumber\\
&&|0101011\rangle+|0100111\rangle+|0011101\rangle+|0011011\rangle+|0010111\rangle+|0001111\rangle+\nonumber\\
&&|1111000\rangle+|1110100\rangle+|1110010\rangle+|1101100\rangle+|1101010\rangle+|1100110\rangle+\nonumber\\
&&|1011100\rangle+|1011010\rangle+|1010110\rangle+|1001110\rangle+|0111100\rangle+|0111010\rangle+\nonumber\\
&&|0110110\rangle+|0101110\rangle+|0011110\rangle+|1111101\rangle+|1111011\rangle+|1110111\rangle+\nonumber\\
&&|1101111\rangle+|1011111\rangle+|0111111\rangle+|1111110\rangle,
\end{eqnarray}
\end{widetext}

is symmetric with respect to permutations and it
has all its two, three, four, five and six qubits reductions identically equal.
That is, every subsystem of $N^{\prime}<7$ qubits contains the same physical information.
Moreover, these reductions have rank two independently of the size of the reduction.
Thus, the above 7-qubits state is a genuine multipartite entangled state but the entanglement of its parties is weak.
 The strong regularity observed in the reductions of this state is due to the high strength $k=6$ of the
 orthogonal array OA(64,7,2,6). That is, every subset of six columns of the OA contains the $2^6$ possible
combinations of the symbols along the 64 rows.

Note that most of the 1--uniform states presented are symmetric under permutation of qubits. It has been proven that a $k$--uniform state is symmetric if $k\leq1$ \cite{Arnaud}. Therefore, $k$--uniform states cannot be symmetric for $k\geq2$. This means that neither 2--uniform the states constructed from Table \ref{Tabla3} nor those appearing in Appendix \ref{AppendixB} are symmetric. Interestingly, this property can be also applied to orthogonal arrays of index unity through Theorem {OAkMM}: an OA of index unity, $N$ runs and $2$ levels is invariants under permutation of columns if it has strength $k=1$. Indeed, given that $r=\lambda 2^k$, an OA of index $\lambda=1$ and $k=1$ have $r=2$ runs. Thus, the only symmetric OA of index unity are one-to-one connected to the GHZ states.

Table \ref{Tabla3} provides a list of some orthogonal arrays of index $\lambda=1$ \cite{Sloane}.
As we have shown along the work, they can be used for a direct construction of $2$--uniform quantum states
of $N$ subsystems, with $d$ levels each:
\begin{table}
\begin{center}
\begin{tabular}{c|c|c|c}
\hspace{0.5cm}Orthogonal array\hspace{0.5cm} & \hspace{0.5cm}$r$\hspace{0.5cm} & \hspace{0.5cm}$N$\hspace{0.5cm} & \hspace{0.5cm}$d$\hspace{0.5cm} \\
   \hline
OA(9,4,3,2)&9&4&3\\
OA(16,5,4,2)&16&5&4\\
OA(25,6,5,2)&25&6&5\\
OA(49,8,7,2)&49&8&7\\
OA(64,9,8,2)&64&9&8\\
OA(81,10,9,2)&81&10&9\\
OA(100,4,10,2)&100&4&10\\
OA(121,12,11,2)&121&12&11\\
OA(144,7,12,2)&144&7&12\\
OA(169,14,13,2)&169&14&13\\
OA(256,17,16,2)&256&17&16\\
OA(289,18,17,2)&289&18&17\\
\end{tabular}
\end{center}
\caption{Orthogonal arrays of index $\lambda=1$ allowing for a direct construction of $2$--uniform states.}
\label{Tabla3}
\end{table}
Note that in most cases the values of the parameter $r$ in the above OA is a power of a  prime, as these arrays
were obtained with Theorem \ref{TeoRao}. However, the cases of OA(64,6,4,3), OA(100,4,10,2) and OA(144,7,12,2)
 have been probably constructed by computing simulations. We remark here
that there is no simple way to generate orthogonal arrays of index $\lambda=1$ for non prime--power values $d$.
However, they might exist. Construction of OA of strength $k=3$ and index $\lambda=1$
is implied by the following theorem \cite{Kounias}:

\begin{teo}[Kounias]
For any even $d$ every orthogonal array OA($d^3,d+1,d,3$) can be extended to OA($d^3,d+2,d,3$).
\end{teo}
This theorem can be restated in the context of multipartite quantum states:
\begin{prop}
For any even $d$ every $3$--uniform state of $d+1$ qudits generated by OA($d^3,d+1,d,3$) can be extended to a $3$--uniform state of
$d+2$ qudits.
\end{prop}
Unfortunately, in current libraries of OA we could not find a suitable example to illustrate this proposition.

\section{Construction of $\mathbf{k}$--uniform states with non-positive terms}\label{AppendixC}
Some orthogonal arrays OA($r,N,d,k$) satisfying $k\leq N/2$ do not lead us to a $k$--uniform state of positive terms. This is because some rows within sets of $N-k$ columns are repeated and, consequently, the reductions to $k$ qudits are not diagonal. However, sometimes is possible to introduce \emph{minus signs} in some terms of the state such that it becomes a $k$--uniform state. Here we carefully illustrate this procedure for a $2$--uniform state of 5 qubits obtained from the Hadamard matrix $H_8$. Despite this \emph{patch} can be applied in many cases we cannot prove the existence of $k$--uniform states of $N$ qudits for every $k$ and $N$ such that $k\leq N/2$. For example, we could not find neither $3$--uniform states of 7 qubits nor $4$--uniform states of 8 qubits. As already mentioned, the problem relies in the violation of the diagonality condition \emph{(B)}.

Let us consider the last 5 columns of the Hadamard matrix $H_8$ given in Eq.(\ref{H8}). Here, $N=5$ and $\kappa=8$ so
inequality (\ref{cotakappa}) does not hold. Let us define the state
\begin{eqnarray}
|\Phi_5'\rangle&=& |11111\rangle+|01010\rangle+|01100\rangle+\nonumber\\
&&|11001\rangle+|10000\rangle+|00101\rangle+\nonumber\\
&&|00011\rangle+|10110\rangle,
\end{eqnarray}
which is not a $2$--uniform state. In order to find a $2$--uniform state we start by adding some unimodular complex numbers in the terms of $|\Phi_5'\rangle$:
\begin{widetext}
\begin{eqnarray}\label{phi53}
|\Phi_5"\rangle(\vec{\alpha})&=& (-1)^{\alpha_0}|11111\rangle+(-1)^{\alpha_1}|01010\rangle+(-1)^{\alpha_2}|01100\rangle+(-1)^{\alpha_3}|11001\rangle+\nonumber\\
&&(-1)^{\alpha_4}|10000\rangle+(-1)^{\alpha_5}|00101\rangle+(-1)^{\alpha_6}|00011\rangle+(-1)^{\alpha_7}|10110\rangle.
\end{eqnarray}
\end{widetext}

So, from every non-maximally mixed reduction we obtain a linear system of equations for the phases $\{\alpha_r\}$. If this system is compatible then there exists states of the form given in Eq.(\ref{phi53}) such that all its reductions are diagonal. Additionally, as we consider orthogonal arrays the reductions are maximally mixed. On the other hand, the linear system of equations is incompatible when $k>N/2$.

For $|\Phi_5'\rangle$ the only non-maximally mixed reductions are $\rho_{34}=\mathrm{Tr}_{125}(\rho_{12345})$ and $\rho_{25}=\mathrm{Tr}_{134}(\rho_{12345})$. In the following three tables:
\begin{center}
\begin{tabular}{rrrrr}
  \hspace{0.2cm}\textbf{4} &\hspace{0.2cm} \textbf{5} & \hspace{0.2cm}\textbf{1} & \hspace{0.2cm}\textbf{2} &\hspace{0.2cm} \textbf{3} \\
  \hline
  1 & 1 &\hspace{0.4cm} 1 & 1 & 1 \\
  1 & 0 & 0 & 1 & 0 \\
  0 & 0 & 0 & 1 & 1 \\
  0 & 1 & 1 & 1 & 0 \\
  0 & 0 & 1 & 0 & 0 \\
  0 & 1 & 0 & 0 & 1 \\
  1 & 1 & 0 & 0 & 0 \\
  1 & 0 & 1 & 0 & 1 \\
\end{tabular}\vspace{1cm}\hspace{0.7cm}
\begin{tabular}{rrrrr}
  \hspace{0.2cm}\textbf{3} &\hspace{0.2cm} \textbf{4} & \hspace{0.2cm}\textbf{1} & \hspace{0.2cm}\textbf{2} &\hspace{0.2cm} \textbf{5} \\
  \hline
  1 & 1 &\hspace{0.4cm} 1 & 1 & 1 \\
  0 & 1 & 0 & 1 & 0 \\
  1 & 0 & 0 & 1 & 0 \\
  0 & 0 & 1 & 1 & 1 \\
  0 & 0 & 1 & 0 & 0 \\
  1 & 0 & 0 & 0 & 1 \\
  0 & 1 & 0 & 0 & 1 \\
  1 & 1 & 1 & 0 & 0 \\
\end{tabular}
\begin{tabular}{rrrrr}
  \hspace{0.2cm}\textbf{2} & \hspace{0.2cm}\textbf{5} & \hspace{0.2cm}\textbf{1} & \hspace{0.2cm}\textbf{3} & \hspace{0.2cm}\textbf{4} \\
  \hline
  1 & 1 &\hspace{0.4cm} 1 & 1 & 1 \\
  1 & 0 & 0 & 0 & 1 \\
  1 & 0 & 0 & 1 & 0 \\
  1 & 1 & 1 & 0 & 0 \\
  0 & 0 & 1 & 0 & 0 \\
  0 & 1 & 0 & 1 & 0 \\
  0 & 1 & 0 & 0 & 1 \\
  0 & 0 & 1 & 1 & 1 \\
\end{tabular}
\end{center}\vspace{0.5cm}
\noindent we can see the orthogonal array version of the process to obtain the reductions $\rho_{45}=\mathrm{Tr}_{123}(\rho_{12345})$, $\rho_{34}=\mathrm{Tr}_{125}(\rho_{12345})$ and $\rho_{25}=\mathrm{Tr}_{134}(\rho_{12345})$. The first table corresponds to the analysis of the reduction $\rho_{45}=\mathbb{I}/4$. Note that the binary numbers determined by the columns \textbf{4} and \textbf{5} cover exactly 2 times the four pairs $\{(0,0);(0,1);(1,0);(1,1)\}$. This happens in every pair of columns because the OA has strength $k=2$. Also, the frequency of appearance of every pair is 2 given that the OA has index $\lambda=2$. Therefore the diagonal of the reduced state $\rho_{45}$ is uniform. Additionally, this reduction is diagonal because the rows determined by the columns \textbf{1}, \textbf{2} and \textbf{3} are different. In the second table we analyze the reduction $\rho_{34}$. Here, the columns \textbf{3} and \textbf{4} determine a uniform diagonal of the reduction. However, the columns \textbf{1},
\textbf{2} and \textbf{5} are repeated, so $\rho_{34}$ is not a diagonal matrix. So is the case for the third table corresponding to $\rho_{25}$). These are the reasons why $|\Phi_5'\rangle$ is not a $2$--uniform state. In what follows, we explain how to modify this construction allowing for negative terms in $|\Phi_5'\rangle$
 to satisfy the $2$--uniform property. In higher dimensions, possibly complex phases are required as $k$--uniform states containing all
  terms with real weights may not exist. Basically, the non-diagonal terms in the reductions arise from an even number of terms.
    If we consider orthogonal arrays all these contributions are positive. Therefore one can change the sign of some terms of
   the state $|\Phi\rangle$ in order to achieve that all the reduced states are diagonal.
Let us concentrate in the middle array in the above table. Making use of the definition
\begin{eqnarray}
\rho_A&=&\mathrm{Tr}_B(\rho_{AB})\nonumber\\
&=&\sum_{i,j,k,l}a_{i,j,k,l}\mathrm{Tr}_B\textbf{(}|i\rangle_A|j\rangle_B{}_A\langle k|{}_B\langle l|\textbf{)}\nonumber\\
&=&\sum_{i,j,k,l}a_{i,j,k,l}\,\langle l|j\rangle\,|i\rangle_A\langle k|,
\end{eqnarray}
and the data from the table above one can construct the following reduction:
\begin{equation}
\rho_{34} = \bordermatrix{~ & \emph{00} & \emph{01} & \emph{10} & \emph{11} \cr
              \emph{00}\hspace{0.3cm} & 1/4 & 0 & 0 & 1/4 \cr
              \emph{01}\hspace{0.3cm} & 0 & 1/4 & 1/4 & 0 \cr
              \emph{10}\hspace{0.3cm} & 0 & 1/4 & 1/4 & 0 \cr
              \emph{11}\hspace{0.3cm} & 1/4 & 0 & 0 & 1/4 \cr}.
\end{equation}
Therefore, assuming that
\begin{widetext}
\begin{eqnarray}
|\Phi_5"\rangle&=& |11111\rangle+(-1)^{\alpha_1}|01010\rangle+(-1)^{\alpha_2}|01100\rangle+(-1)^{\alpha_3}|11001\rangle+\nonumber\\
&&(-1)^{\alpha_4}|10000\rangle+(-1)^{\alpha_5}|00101\rangle+(-1)^{\alpha_6}|00011\rangle+(-1)^{\alpha_7}|10110\rangle,
\end{eqnarray}
\end{widetext}
we have the following linear system of equations:
\begin{eqnarray}
\alpha_0+\alpha_3+\alpha_4+\alpha_7+1=0,\nonumber\\
\alpha_1+\alpha_2+\alpha_5+\alpha_6+1=0.
\end{eqnarray}
Interestingly, the same linear equations appear for the reduction $\rho_{25}$. For example, a solution arises from considering that the only non--null phases are $\alpha_6=\alpha_7=1$. Consequently, the state:
\begin{eqnarray}
|\Phi_5"\rangle&=& |11111\rangle+|01010\rangle+|01100\rangle+\nonumber\\
&&|11001\rangle+|10000\rangle+|00101\rangle-\nonumber\\
&&|00011\rangle-|10110\rangle,
\end{eqnarray}
is a $2$--uniform state of $N=5$ qubits consisting of eight terms, out of which two are negative.

\end{document}